\documentclass{ieeeaccess}

\newcommand{\forarxiv}{0}
\usepackage[utf8]{inputenc}
\usepackage{amsmath}
\usepackage{amsthm}
\usepackage{amssymb}
\usepackage{graphicx}
\usepackage{bm}
\usepackage{dsfont}
\usepackage{subcaption}
\usepackage{tikz}

\usepackage[sort,compress]{cite}
\usepackage{amsfonts}
\usepackage{url}
\usepackage[pdftex, pdfstartview={FitV}, pdfpagelayout={TwoColumnLeft},bookmarksopen=true,plainpages = false, colorlinks=true, linkcolor=black, citecolor = black, urlcolor = blue,filecolor=black , pagebackref=false,hypertexnames=false, plainpages=false, pdfpagelabels ]{hyperref}
\usepackage{balance}
\usepackage[capitalize]{cleveref}

\newtheorem{theorem}{Theorem}[section]
\newtheorem{corollary}[theorem]{Corollary}
\newtheorem{lemma}[theorem]{Lemma}

\crefname{section}{Section}{Sections}
\crefname{theorem}{Theorem}{Theorems}
\crefname{lemma}{Lemma}{Lemmas}
\crefname{table}{Table}{Tables}
\crefformat{equation}{(#2#1#3)}
\crefname{algocf}{Algorithm}{Algorithms}
\Crefname{algocf}{Algorithm}{Algorithms}
\crefname{ALC@unique}{Line}{Lines}

\newcommand{\Nu}{\mathbf{n}^U}
\newcommand{\Mu}{\mathbf{M}^U}

\newcommand{\Aout}{\mathbf{A}^\text{out}}
\newcommand{\bout}{\mathbf{b}^\text{out}}
\newcommand{\Ain}{\mathbf{A}^\text{in}}

\newcommand{\x}{\mathbf{x}}

\newcommand{\y}{\mathbf{y}}

\newcommand{\R}{\mathds{R}}

\newcommand{\istate}{k}
\newcommand{\ifacet}{i}
\newcommand{\icontrol}{j}
\newcommand{\ilayer}{l}

\newcommand{\piclbu}{\bm{\Gamma}}
\newcommand{\piclbl}{\bm{\Delta}}
\newcommand{\piclAu}{\bm{\Upsilon}}
\newcommand{\piclAl}{\bm{\Xi}}

\newcommand{\mout}{m_\text{out}}

\newcommand{\Xt}{\mathcal{X}_{t}}

\newcommand{\Xtt}{\mathcal{X}_{t+1}}

\newcommand{\xinX}{\mathbf{x}_t \in \Xt}

\newcommand{\piUCL}{\pi^{U_{CL}}_{:,\ifacet}}
\newcommand{\piLCL}{\pi^{L_{CL}}_{:,\ifacet}}
\newcommand{\Nuj}{\mathbf{n}^U_{\ifacet}}
\newcommand{\Muj}{\mathbf{M}^U_{\ifacet,:}}
\newcommand{\Nlj}{\mathbf{n}^L_{\ifacet}}
\newcommand{\Mlj}{\mathbf{M}^L_{\ifacet,:}}

\newcommand{\Ainbin}{\mathbf{A}^\text{in}\mathbf{x}_t\leq \mathbf{b}^\text{in}}

\newcommand{\Aoutj}{\mathbf{A}_{\ifacet,:}^{\textrm{out}}}
\newcommand{\AoutjBt}{\Aoutj \mathbf{B}_{t}}
\newcommand{\AoutjBtj}{\Aoutj \mathbf{B}_{t,:,\icontrol}}
\newcommand{\AoutjBtPiclAuCt}{\Aoutj \mathbf{B}_{t} \piclAu_{\ifacet,:,:} \mathbf{C}_t^T}
\newcommand{\AoutjBtPiclAlCt}{\Aoutj \mathbf{B}_{t} \piclAl_{\ifacet,:,:} \mathbf{C}_t^T}

\newcommand{\boutj}{\bout_{\ifacet}}

\newcommand{\xnom}{\mathring{\mathbf{x}}}
\newcommand{\ynom}{\mathring{\mathbf{y}}}
\newcommand{\xtnom}{\xnom_{t}}
\newcommand{\bpxnomscalar}{\mathcal{B}_p(\xnom, \epsilon)}
\newcommand{\bpxnom}{\mathcal{B}_p(\xnom, \bm{\epsilon)}}
\newcommand{\bpynom}{\mathcal{B}_p(\ynom, \epsilon)}
\newcommand{\binfxnom}{\mathcal{B}_{\infty}(\xnom, \epsilon)}
\newcommand{\bpxtnom}{\mathcal{B}_p(\xtnom, \epsilon)}
\newcommand{\xinXball}{\mathbf{x}_t \in \bpxtnom}

\newcommand{\xttbnd}{\bm{\gamma}}

\newcommand{\xttubistate}{\xttbnd^U_{t+1,\istate}}
\newcommand{\xttlbistate}{\xttbnd^L_{t+1,\istate}}
\newcommand{\xttubifacet}{\xttbnd^U_{t+1,\ifacet}}
\newcommand{\xttlbifacet}{\xttbnd^L_{t+1,\ifacet}}
\newcommand{\xttub}{\xttbnd^U_{t+1}}
\newcommand{\xttlb}{\xttbnd^L_{t+1}}

\newcommand{\xttlbonelp}{\xttbnd^L_{1,\istate}}
\newcommand{\xttubonelp}{\xttbnd^U_{1,\istate}}
\newcommand{\xttlbonepoly}{\xttbnd^L_{1,\ifacet}}
\newcommand{\xttubonepoly}{\xttbnd^U_{1,\ifacet}}

\newcommand{\CROWNAu}{\mathbf{\Psi}}
\newcommand{\CROWNAl}{\mathbf{\Phi}}
\newcommand{\CROWNbu}{\bm{\alpha}}
\newcommand{\CROWNbl}{\bm{\beta}}

\newcommand{\backprojA}{
\begin{bmatrix}
\mathbf{A}_{t} + \mathbf{Z}^{\mathbf{B}\CROWNAu\CROWNAl} \\
- \left(\mathbf{A}_{t} + \mathbf{Z}^{\mathbf{B}\CROWNAl\CROWNAu}\right) \\
\mathbf{I}_{n_x} \\
- \mathbf{I}_{n_x}
\end{bmatrix}}
\newcommand{\backprojb}{
\begin{bmatrix}
\bar{\mathbf{x}}_{t+1} - \left( \mathbf{z}^{\mathbf{B}\CROWNbu\CROWNbl} + \mathbf{c}_t \right) \\
- \ubar{\mathbf{x}}_{t+1} + \mathbf{z}^{\mathbf{B}\CROWNbl\CROWNbu} + \mathbf{c}_t \\
\bar{\mathbf{x}}_{t}\\
-\ubar{\mathbf{x}}_{t}
\end{bmatrix}
}

\newcommand{\Jbar}[3]{s(#1, #2, #3)}

\usepackage{accents}
\newcommand{\ubar}[1]{\underaccent{\bar}{#1}}

\DeclareMathOperator*{\argmax}{\arg\!\max}

\DeclareMathOperator{\Tr}{Tr}
\crefformat{chapter}{\S#2#1#3}
\crefmultiformat{chapter}{\S\S#2#1#3}{and~#2#1#3}{, #2#1#3}{, and~#2#1#3}
\crefformat{section}{\S#2#1#3}
\crefmultiformat{section}{\S\S#2#1#3}{and~#2#1#3}{, #2#1#3}{, and~#2#1#3}

\usepackage{algorithm,algorithmic}

\newcommand{\shortleftarrow}[1]{%
\parbox{#1}{\tikz{\draw[<-](0,0)--(#1,0);}}
}

\NewSpotColorSpace{PANTONE}
\AddSpotColor{PANTONE} {PANTONE3015C} {PANTONE\SpotSpace 3015\SpotSpace C} {1 0.3 0 0.2}
\SetPageColorSpace{PANTONE}%
\definecolor{accessblue}{RGB}{0,98,155}
\definecolor{greycolor}{cmyk}{0,0,0,.8}
\definecolor{grey}{cmyk}{0,0,0,.1}
\definecolor{black}{cmyk}{0,0,0,1}

\begin{document}

\history{Date of publication Dec 6, 2021, date of current version Dec 6, 2021 (this is the accepted version of the article).}
\doi{https://doi.org/10.1109/ACCESS.2021.3133370}

\newcommand{\papertitle}{Reachability Analysis of Neural Feedback Loops}

\title{\papertitle}

\author{
\if \forarxiv 1
Michael~Everett,
Golnaz Habibi,
Chuangchuang Sun,
and~Jonathan~P.~How% <-this % stops a space
\else
\uppercase{Michael~Everett},~\IEEEmembership{Member,~IEEE},
\uppercase{Golnaz Habibi},~\IEEEmembership{Member,~IEEE},
\uppercase{Chuangchuang Sun},~\IEEEmembership{Member,~IEEE},
\uppercase{and~Jonathan~P.~How},~\IEEEmembership{Fellow,~IEEE}% <-this % stops a space
\fi
\address[]{Authors are with the Aerospace Controls Laboratory, Massachusetts Institute of Technology, Cambridge,
MA, 02139 USA}% <-this % stops a space
\tfootnote{This work is funded in part by Ford Motor Company, by ARL DCIST under Cooperative Agreement Number W911NF-17-2-0181, and by Scientific Systems Company, Inc. under research agreement \# SC-1661-04. \\ \textbf{Code:} \url{https://github.com/mit-acl/nn_robustness_analysis}}
}

% The paper headers
\if \forarxiv 1
\markboth{(In Review)}%
{Everett \MakeLowercase{\textit{et al.}}: \papertitle}
\else
\markboth
{Everett \headeretal: \papertitle}
{Everett \headeretal: \papertitle}
\corresp{Corresponding author: Michael Everett (e-mail: mfe@mit.edu).}
\fi

%%%%%%%%%%%%%%%%%%%%%%%%%%%%%%%%%%%%%%%%%%%%%%%%%%%%%%%%%%%%%%%%%%%%%%%%%%%%%%%%
\begin{abstract}
Neural Networks (NNs) can provide major empirical performance improvements for closed-loop systems, but they also introduce challenges in formally analyzing those systems' safety properties.
In particular, this work focuses on estimating the forward reachable set of \textit{neural feedback loops} (closed-loop systems with NN controllers).
Recent work provides bounds on these reachable sets, but the computationally tractable approaches yield overly conservative bounds (thus cannot be used to verify useful properties), and the methods that yield tighter bounds are too intensive for online computation.
This work bridges the gap by formulating a convex optimization problem for the reachability analysis of closed-loop systems with NN controllers.
While the solutions are less tight than previous (semidefinite program-based) methods, they are substantially faster to compute, and some of those computational time savings can be used to refine the bounds through new input set partitioning techniques, which is shown to dramatically reduce the tightness gap.
The new framework is developed for systems with uncertainty (e.g., measurement and process noise) and nonlinearities (e.g., polynomial dynamics), and thus is shown to be applicable to real-world systems.
To inform the design of an initial state set when only the target state set is known/specified, a novel algorithm for backward reachability analysis is also provided, which computes the set of states that are guaranteed to lead to the target set.
The numerical experiments show that our approach (based on linear relaxations and partitioning) gives a $5\times$ reduction in conservatism in $150\times$ less computation time compared to the state-of-the-art.
Furthermore, experiments on quadrotor, 270-state, and polynomial systems demonstrate the method's ability to handle uncertainty sources, high dimensionality, and nonlinear dynamics, respectively.
\end{abstract}

% Note that keywords are not normally used for peerreview papers.

\titlepgskip=-15pt

\maketitle

%!TEX root=main.tex

\section{Introduction}\label{sec:intro}

\begin{figure}
	\centering
	\includegraphics[page=1,width=0.8\linewidth,trim=250 150 240 70,clip]{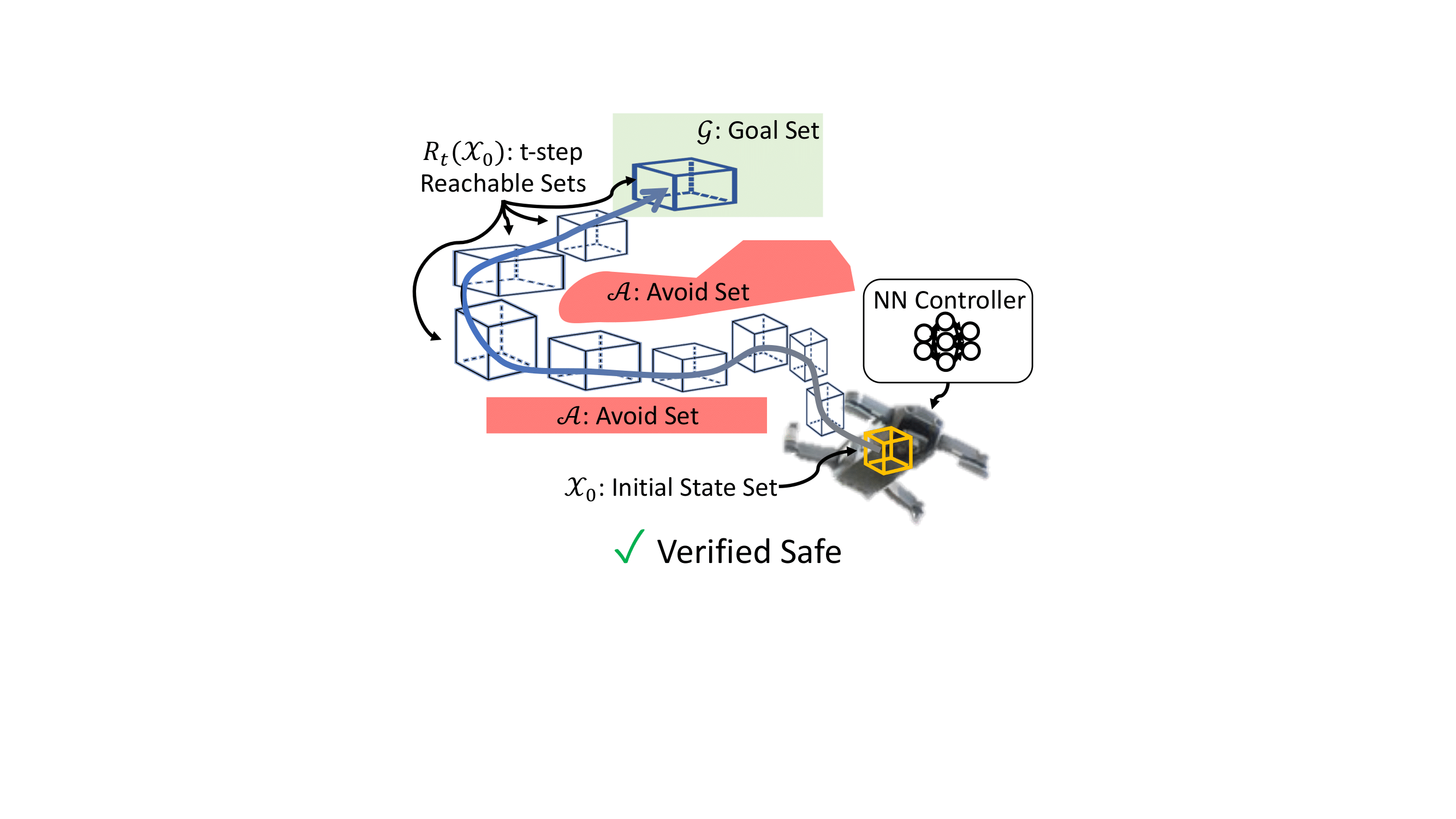}
	\caption{Forward Reachability Analysis. The objective is to compute blue sets $\mathcal{R}_t(\mathcal{X}_0)$, to ensure a system starting in $\mathcal{X}_0$ (yellow) ends in $\mathcal{G}$ (green) and avoids $\mathcal{A}_0,\mathcal{A}_1$ (red). This is challenging for systems with NN control policies.}
	\label{fig:problem_cartoon}
	\vspace{-0.2in}
\end{figure}

\IEEEPARstart{N}{eural} Networks (NNs) are pervasive in many fields because of their ability to express highly general input-output relationships, such as for perception, planning, and control tasks in robotics.
However, before deploying NNs on safety-critical systems, there must be techniques to guarantee that the closed-loop behavior of systems with NNs, which we call \textit{neural feedback loops} (NFLs), will meet desired specifications.
The goal of this paper is to develop a framework for guaranteeing that systems with NN controllers will reach their goal states while avoiding undesirable regions of the state space, as in~\cref{fig:problem_cartoon}.

Despite the importance of analyzing closed-loop behavior, much of the recent work on formal NN analysis has focused on NNs in isolation (e.g., for image classification)~\cite{Ehlers_2017, Katz_2017, Huang_2017b,Lomuscio_2017,Tjeng_2019,Gehr_2018}, with an emphasis on efficiently relaxing NN nonlinearities~\cite{gowal2018effectiveness,Weng_2018,singh2018fast,zhang2018efficient,Wong_2018,Raghunathan_2018,fazlyab2019safety}.
On the other hand, closed-loop system reachability has been studied for decades, but traditional approaches, such as Hamilton-Jacobi methods~\cite{tomlin2000game,bansal2017hamilton}, do not consider NNs in the loop.

Several recent works~\cite{dutta2019reachability,huang2019reachnn,fan2020reachnn,ivanov2019verisig,xiang2020reachable,hu2020reach} propose methods that compute forward reachable sets of NFLs.
A key challenge is in maintaining computational efficiency while still providing tight bounds on the reachable sets.
In addition, large initial state sets present challenges as the underlying NN relaxations become loose, leading to loose reachable set estimates.
The literature also typically assumes perfect knowledge of  system dynamics, with no stochasticity.
Furthermore, the literature focuses on forward reachability analysis (what states could the system end in, starting from a given $\mathcal{X}_0$?), but backward reachability analysis (what states can the system start from, to end in a given $\mathcal{X}_T$?) is an equally important problem in safety analysis.
While forward and backward reachability differ only by a change of variables in classical systems~\cite{bansal2017hamilton,evans1998partial}, both the nonlinearities and matrix dimensions of NN controllers raise fundamental challenges that prevent propagating sets backward through a NFL.

% To create a scalable reachability analysis framework, we introduce a linear programming-based framework that leverages tools from~\cite{zhang2018efficient} to relax the NN's nonlinearities.
% While this relaxation provides substantial improvement in computational efficiency over a state-of-the-art method that uses semi-definite programming, the linear relaxations also introduce some conservatism.
% Thus, the proposed algorithm trades off some computational efficiency for bound tightness by partitioning the input set, as motivated by~\cite{xiang2018reachable,xiang2020reachable,everett2020robustness}.
% Finally, the proposed framework considers measurement, process noise, and can be extended to handle polynomial dynamics, thus being more amenable to applications on real, uncertain closed-loop systems.

To address these challenges, this work's contributions include:
\begin{itemize}
\item A linear programming-based formulation of forward reachability analysis for NFLs, which provides a computationally efficient method for verifying safety properties,
\item The use of input set partitioning techniques to provide tight bounds on the reachable sets despite large initial state sets,
\item The consideration of measurement noise, process noise, and nonlinear dynamics, which improves the applicability to real systems with uncertainty, and
\item A framework for backward reachability analysis of NFLs, which enables estimating which starting states will lead to a target set despite non-invertible NN weight matrices and activations.
\end{itemize}
Numerical experiments show that the new method provides $5\times$ better accuracy in $150\times$ less computation time  compared to~\cite{hu2020reach} and that the method can handle uncertainty sources, high dimensionality, and nonlinear dynamics via applications on quadrotor, 270-state, and polynomial systems.

This article extends our conference paper~\cite{Everett21_ICRA} by providing an improved problem formulation, notation, and algorithmic details. It also includes:
\begin{itemize}
    \item A closed-form solution to provide an order-of-magnitude computational speedup for the same bounds in~\cref{sec:forward_reachability:closed_form_soln},
    \item An extension of state-of-the-art partitioning techniques to the NFL setting in \cref{sec:partition},
    \item Formulations that handle nonlinearities (control limits and polynomial terms) in the dynamics in \cref{sec:nonlinear},
    \item A framework for backward reachability analysis in \cref{sec:backward_reachability}.
\end{itemize}

Open-source software implementations of this paper's algorithms and results can be found at \url{https://github.com/mit-acl/nn_robustness_analysis}.

%!TEX root=main.tex

\section{Related Work}\label{sec:related_work}
Related work on reachability analysis can be categorized into works on NNs in isolation, closed-loop systems without NNs, and closed-loop systems with NNs.
For instance, machine learning literature includes many methods to verify properties of NNs, often motivated by defending against adversarial examples~\cite{Szegedy_2014}.
These methods broadly range from exact~\cite{katz2017reluplex} to tight~\cite{fazlyab2019safety} to efficient~\cite{zhang2018efficient} to fast~\cite{gowal2018effectiveness}.
Although these tools are not designed for closed-loop systems, the  NN relaxations from \cite{zhang2018efficient} provide a key foundation here.

For closed-loop systems, reachability analysis is a standard component of safety verification.
Modern methods include Hamilton-Jacobi Reachability methods~\cite{tomlin2000game,bansal2017hamilton}, SpaceEx~\cite{frehse2011spaceex}, Flow*~\cite{chen2013flow}, CORA~\cite{althoff2015introduction}, and C2E2~\cite{duggirala2015c2e2,fan2016automatic}, but these do not account for NN control policies.
Orthogonal approaches that do not explicitly estimate the system's forward reachable set, but provide other notions of safety, include Lyapunov function search~\cite{papachristodoulou2002construction} and control barrier functions (CBFs)~\cite{ames2016control}.

Recent reachability analysis approaches that do account for NN control policies face a tradeoff between computation time and conservatism.
\cite{dutta2019reachability,huang2019reachnn,fan2020reachnn} use polynomial approximations of NNs to make the analysis tractable.
Most works consider NNs with ReLU approximations, whereas \cite{ivanov2019verisig} considers sigmoidal activations.
\cite{xiang2020reachable,yang2019efficient} introduce conservatism by assuming the NN controller could output its extreme values at every state.
Most recently, \cite{hu2020reach} formulated the problem as a SDP, called Reach-SDP.
This work builds on both~\cite{xiang2020reachable,hu2020reach} and makes the latter more scalable by re-formulating the SDP as a linear program, introduces sources of uncertainty in the closed-loop dynamics, and shows further improvements by partitioning the input set.

%!TEX root=main.tex

\section{Preliminaries}

\subsection{Closed-Loop System Dynamics}

\begin{figure}[t]
    \centering
    \includegraphics[page=4,width=\linewidth,trim=100 100 100 100,clip]{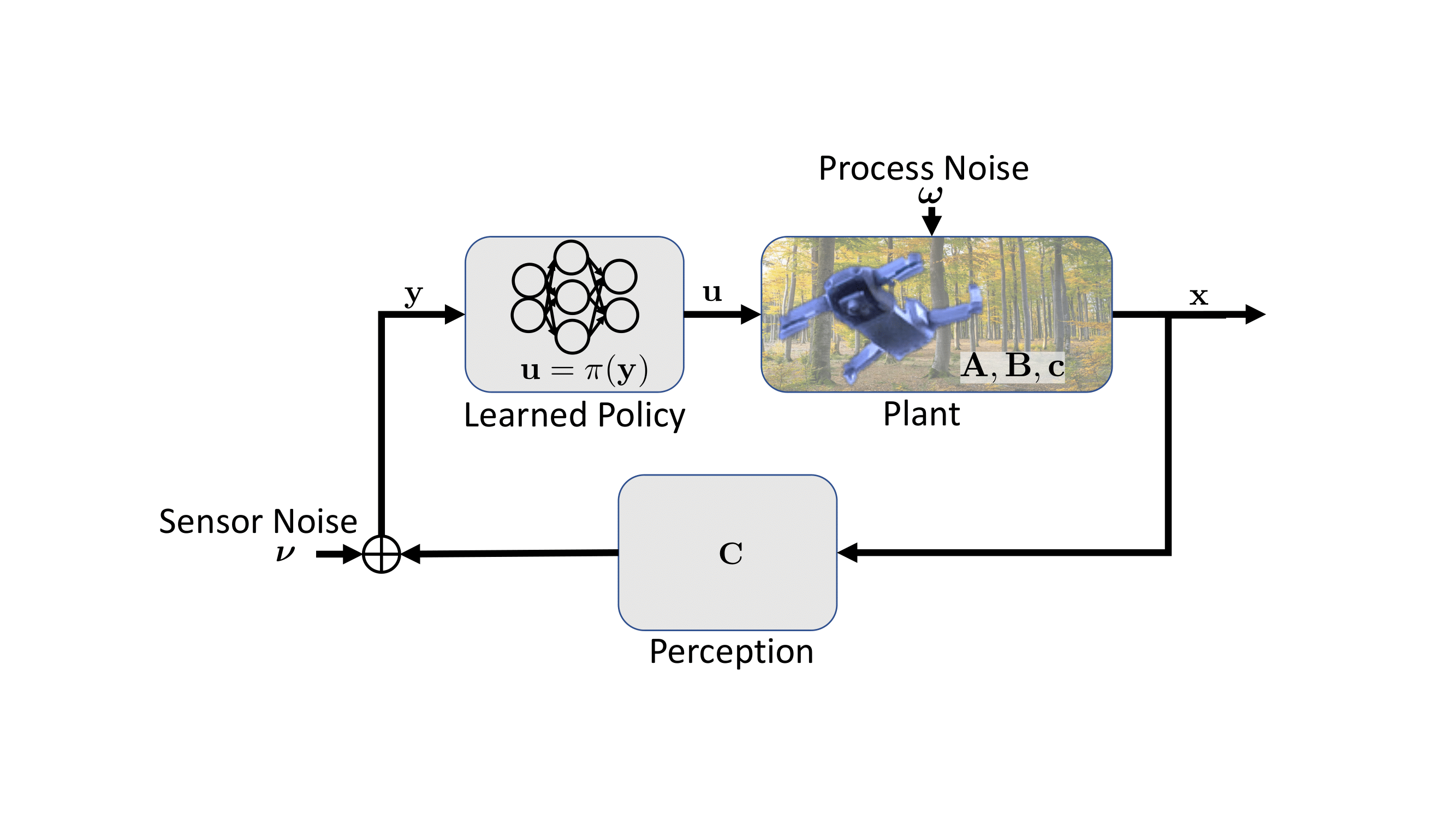}
    \caption{Neural Feedback Loop. Many real world deployments of NNs are part of complex, closed-loop systems (e.g., robots with learned policies take actions that modify the world state and influence future observations).
    }
    \label{fig:neural_feedback_loop}
\end{figure}

Consider a discrete-time linear time-varying system,
\begin{align}
\begin{split}
    \mathbf{x}_{t+1} &= \mathbf{A}_t \mathbf{x}_{t} + \mathbf{B}_t \mathbf{u}_t + \mathbf{c}_t + \bm{\omega}_t \label{eqn:ltv_dynamics} \\
    \mathbf{y}_{t} &= \mathbf{C}_t^T\mathbf{x}_t + \bm{\nu}_t,
\end{split}
\end{align}
where $\mathbf{x}_t\in\R^{n_x},\, \mathbf{u}_t\in\R^{n_u},\, \mathbf{y}_t\in\R^{n_y}$ are state, control, and output vectors, $\mathbf{A}_t,\, \mathbf{B}_t,\, \mathbf{C}_t$ are known system matrices, $\mathbf{c}_t\in\R^{n_x}$ is a known exogenous input, and $\bm{\omega}_t\sim\Omega$ and $\bm{\nu}_t\sim N$ are process and measurement noises sampled at each timestep from unknown distributions with known, finite support (i.e., $\bm{\omega}_t\in [\ubar{\bm{\omega}}_t,\, \bar{\bm{\omega}}_t ], \bm{\nu}_t\in [\ubar{\bm{\nu}}_t,\, \bar{\bm{\nu}}_t ]$ element-wise).

We assume an output-feedback controller $\mathbf{u}_t=\pi(\mathbf{y}_t)$ parameterized by an $m$-layer feed-forward NN, optionally subject to control constraints, $\mathbf{u}_t\in\mathcal{U}_t$.
We denote the closed-loop system with dynamics~\cref{eqn:ltv_dynamics} and control policy $\pi$ as
\begin{equation}
    \mathbf{x}_{t+1} = f(\mathbf{x}_{t}; \pi). \label{eqn:closed_loop_dynamics}
\end{equation}

This NFL is visualized in~\cref{fig:neural_feedback_loop}.
At each timestep, the state $\mathbf{x}$ enters the perception block and is perturbed by sensor noise $\bm{\nu}$ to create observation $\mathbf{y}$.
The observation is fed into the learned policy, which is a NN that selects the control input $\mathbf{u}$.
The control input enters the plant, which is defined by $\mathbf{A}, \mathbf{B}, \mathbf{c}$ matrices and subject to process noise $\bm{\omega}$.
The output of the plant is the state vector at the next timestep.

\subsection{Reachable Sets}

For the closed-loop system~\cref{eqn:closed_loop_dynamics}, we denote $\mathcal{R}_t(\mathcal{X}_0)$ the forward reachable set at time $t$ from a given set of initial conditions $\mathcal{X}_0\subseteq\R^{n_x}$, which is defined by the recursion
\begin{equation}
    \mathcal{R}_{t+1}(\mathcal{X}_0) = f(\mathcal{R}_t(\mathcal{X}_0); \pi), \qquad \mathcal{R}_0(\mathcal{X}_0)=\mathcal{X}_0.\label{eqn:reachable_sets}
\end{equation}

\subsection{Finite-Time Reach-Avoid Verification Problem}\label{sec:background:reach_avoid_verification_problem}

The finite-time reach-avoid property verification problem is defined as follows: Given a goal set $\mathcal{G}\subseteq\R^{n_x}$, a sequence of avoid sets $\mathcal{A}_t\subseteq\R^{n_x}$, and a sequence of reachable set estimates $\mathcal{R}_t\subseteq\R^{n_x}$, determining that every state in the final estimated reachable set will be in the goal set and any state in the estimated reachable sets will not enter an avoid set requires computing set intersections, $\texttt{VERIFIED}(\mathcal{G},\mathcal{A}_{0:N},\mathcal{R}_{0:N})\equiv{\mathcal{R}_N\subseteq\mathcal{G}}\ \mathrm{\&}\ {\mathcal{R}_t\cap\mathcal{A}_t=\emptyset}, \forall t\in\{0,\ldots,N\}$.
% Given an estimate of the reachable sets, it is straightforward to verify finite-time reach-avoid properties.
% That is, given a goal set $\mathcal{G}\subseteq\R^{n_x}$, a sequence of avoid sets $\mathcal{A}_t\subseteq\R^{n_x}$, and a sequence of reachable set estimates $\mathcal{R}_t\subseteq\R^{n_x}$, determining that every state in the final estimated reachable set will be in the goal set and any state in the estimated reachable sets will not enter an avoid set is equivalent to computing set intersections:
% \begin{align}
% &\texttt{VERIFIED}(\mathcal{G},\mathcal{A}_0,\ldots,\mathcal{A}_N,\mathcal{R}_0,\ldots,\mathcal{R}_N)= \label{eqn:defn_verified}\\
%     &\begin{cases}
%         \mathrm{True}, & \mathrm{if}\ \mathcal{R}_N\subseteq\mathcal{G}\ \mathrm{\&}\ \mathcal{R}_t\cap\mathcal{A}_t=\emptyset, \forall t\in\{0,\ldots,N\} \\
%         \mathrm{False}, & \mathrm{otherwise}
%     \end{cases}. \nonumber
% \end{align}

In the case of our nonlinear closed-loop system~\cref{eqn:closed_loop_dynamics}, where computing the reachable sets exactly is computationally intractable, we can instead compute outer-approximations of the reachable sets, $\bar{\mathcal{R}}_t(\mathcal{X}_0)\supseteq\mathcal{R}_t(\mathcal{X}_0)$.
This is useful if the finite-time reach-avoid properties of the system as described by outer-approximations of the reachable sets are verified, because that implies the finite-time reach-avoid properties of the \textit{exact} closed loop system are verified as well.
Tight outer-approximations of the reachable sets are desirable, as they enable verification of tight goal and avoid set specifications, and they reduce the chances of verification being unsuccessful even if the exact system meets the specifications.

\subsection{Control Policy Neural Network Structure}

\newcommand{\NNOperator}{\mathbf{F}}

Using notation from~\cite{zhang2018efficient}, for the $m$-layer neural network used in the control policy, the number of neurons in each layer is $n_\ilayer,\ \forall \ilayer \in [m]$, where $[a]$ denotes the set $\{1,2,\ldots,a\}$.
Let the $\ilayer$-th layer weight matrix be $\mathbf{W}^{(\ilayer)}\in\R^{n_\ilayer\times n_{\ilayer-1}}$ and bias vector be $\mathbf{b}^{(\ilayer)}\in\R^{n_\ilayer}$, and let $\NNOperator_\ilayer: \R^{n_x}\to\R^{n_\ilayer}$ be the operator mapping from network input (measured output vector $\mathbf{y}_t$) to layer $\ilayer$.
We have $\NNOperator_\ilayer(\mathbf{y}_t)=\sigma(\mathbf{W}^{(\ilayer)}\NNOperator_{\ilayer-1}(\mathbf{y}_t)+\mathbf{b}^{(\ilayer)}),\forall \ilayer\in [m-1]$, where $\sigma(\cdot)$ is the coordinate-wise activation function.
The framework applies to general activations, including ReLU, $\sigma(\mathbf{z})=\mathrm{max}(0,\mathbf{z})$.
The network input $\NNOperator_0(\mathbf{y}_t)=\mathbf{y}_t$ produces the (unclipped) control input,
\begin{equation}
    \mathbf{u}_t=\pi(\mathbf{y}_t)=\NNOperator_m(\mathbf{y}_t)=\mathbf{W}^{(m)}\NNOperator_{m-1}(\mathbf{y}_t)+\mathbf{b}^{(m)}.
\end{equation}

\subsection{Neural Network Robustness Verification}

A key step in quickly computing reachable sets of the NFL~\cref{eqn:closed_loop_dynamics} is to relax the nonlinear constraints induced by the NN's nonlinear activation functions.
The relaxation converts each nonlinearity into a linear upper and lower bound, where each bound holds within the known range of inputs to the activation.

To represent the range of inputs to the activation functions, we first define the $\epsilon$-ball.
Denote the $\epsilon$-ball (also called the $\ell_p$-ball) under the $\ell_p$ norm, centered at $\xnom$, with scalar radius, $\epsilon$,
\begin{align}
    \bpxnomscalar &= \{\mathbf{x}\ \lvert\ \lvert\lvert \mathbf{x} - \xnom \rvert \rvert_p \leq \epsilon \}.
\end{align}
The $\epsilon$-ball is extended to the case of vector $\bm{\epsilon}\in\R^n_{\geq 0}$ (i.e., $\bm{\epsilon}$-ball), defined as
\begin{align}
    \bpxnom &= \{\mathbf{x}\ \lvert\ \lim_{\bm{\epsilon}' \to \bm{\epsilon}^+} \lvert\lvert (\mathbf{x} - \xnom) \oslash \bm{\epsilon}' \rvert\rvert_p \leq 1\},
\end{align}
where $\oslash$ denotes element-wise division.

% Given an $m$-layer neural network function $f : \R^{n_0} \rightarrow \R^{n_m}$, there exists two explicit functions $f^L_j : \R^{n_0} \rightarrow \R$ and $f^U_j :\R^{n_0} \rightarrow \R$ such that $\forall j \in [n_m], \; \forall \x \in \Ball$, the inequality $\, f_{j}^{L}(\x) \leq f_{j}(\x) \leq f_{j}^{U}(\x)$ holds true, where
\begin{theorem}[From~\cite{zhang2018efficient}, Convex Relaxation of NN]\label{thm:crown_particular_x}
Given an $m$-layer neural network control policy $\pi:\R^{n_y}\to\R^{n_u}$, there exist two explicit functions $\pi_j^L: \R^{n_y}\to\R^{n_u}$ and $\pi_j^U: \R^{n_y}\to\R^{n_u}$ such that $\forall j\in [n_m], \forall \mathbf{y}\in\bpynom$, the inequality $\pi_j^L(\mathbf{y})\leq \pi_j(\mathbf{y})\leq \pi_j^U(\mathbf{y})$ holds true, where
\begin{align}
\label{eq:f_j_UL}
    \pi_{j}^{U}(\y) &= \CROWNAu_{j,:} \y + \CROWNbu_j \\
    \pi_{j}^{L}(\y) &= \CROWNAl_{j,:} \y + \CROWNbl_j,
    % \pi_{j}^{U}(\y) &= \Au{(0)}_{j,:} \y + \sum_{k=1}^{m}\Au{(k)}_{j,:}(\bias{(k)}+\upbias{(k)}_{:,j}) \\
    % \pi_{j}^{L}(\y) &= \Al{(0)}_{j,:} \y + \sum_{k=1}^{m}\Al{(k)}_{j,:}(\bias{(k)}+\lwbias{(k)}_{:,j}),
\end{align}
where $\CROWNAu, \CROWNAl \in \R^{n_u \times n_y}$ and $\CROWNbu, \CROWNbl \in \R^{n_u}$ are defined recursively using NN weights, biases, and activations (e.g., ReLU, sigmoid, tanh), as detailed in~\cite{zhang2018efficient}.
\end{theorem}

In a closed-loop system, \cref{thm:crown_particular_x} bounds the control output for a \textit{particular} measurement $\mathbf{y}$.
Moreover, if all that is known is $\mathbf{y}\in\bpynom$, \cref{thm:crown_particular_x} provides affine relationships between $\mathbf{y}$ and $\mathbf{u}$ (i.e., bounds valid within the known set of possible $\y$).
These relationships enable efficient calculation of NN output bounds, using Corollary 3.3 of~\cite{zhang2018efficient}.

We could leverage~\cite{zhang2018efficient} to compute reachable sets by first bounding the possible controls, then bounding the next state set by applying the extreme controls from each state.
This is roughly the approach in~\cite{xiang2020reachable,yang2019efficient}, for example.
However, this introduces excessive conservatism, because both extremes of control would not be applied at every state (barring pathological examples).
To produce tight bounds on the reachable sets, we leverage the relationship between measured output and control in~\cref{sec:forward_reachability}.

%!TEX root=main.tex

\section{Forward Reachability Analysis}\label{sec:forward_reachability}

\begin{figure*}[t]
    \centering
    \includegraphics[page=2,width=0.8\linewidth,trim=0 170 0 110,clip]{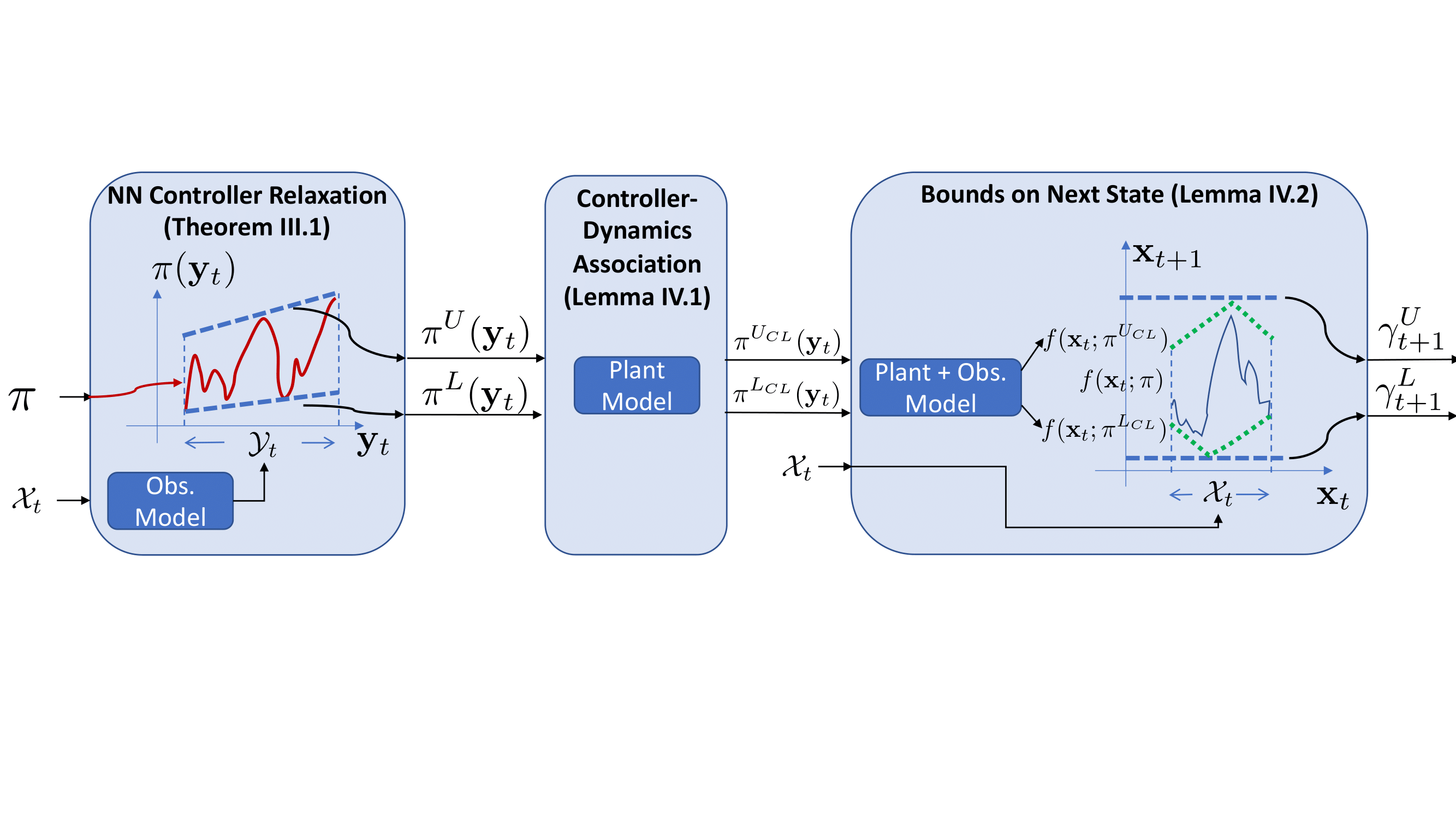}
    \caption{Approach Overview for simple 1D system. \cref{thm:crown_particular_x} relaxes the NN to give affine relationships between observation $\y_t$ and control: $\pi^U, \pi^L$. \cref{thm:particular_xt} uses the system dynamics to associate $\pi^U, \pi^L$ with the next state set. \cref{thm:bounds_on_xt1} optimizes the closed-loop dynamics over all states $\xinX$ to compute bounds on the next state, $\xttub, \xttlb$.}
    \label{fig:approach_flow}
    \vspace{-0.2in}
\end{figure*}

Recall that our goal is to find the set of all possible next states, $\mathbf{x}_{t+1}\in\Xtt$, given that the current state lies within a known set, $\mathbf{x}_t \in \Xt$.
This will allow us to compute reachable sets recursively starting from an initial set $\Xt=\mathcal{X}_0$.

The approach follows the architecture in~\cref{fig:approach_flow}.
After first relaxing the NN controller using~\cref{thm:crown_particular_x}, we then associate linearized extreme controllers with extreme next states in~\cref{sec:forward_reachability:particular_xt}.
Then, using the linearized extreme controller, we optimize over all states in the input set to find extreme next states in~\cref{sec:forward_reachability:bounds_xt1}.
We extend the formulation to handle control limits in~\cref{sec:forward_reachability:control_limits}, then describe how to convert the solutions of the optimization problems into reachable set descriptions in~\cref{sec:forward_reachability:reachable_sets}.

\subsection{Assumptions \& Problem Statement}
This work assumes that $\Xt$ is described by either:
\begin{itemize}
	\item an $\ell_p$-ball for some norm $p\in[1,\infty]$, radius $\bm{\epsilon$}, and centroid $\xnom$, s.t. $\Xt=\bpxnom$; or
	\item a polytope, for some $\Ain\in\R^{m_\text{in}\times n_x}$, $\mathbf{b}^\text{in}\in\R^{m_\text{in}}$, s.t. $\Xt=\{\mathbf{x}_t\ \lvert\ \Ainbin\}$,
\end{itemize}
and shows how to compute $\Xtt$ as described by either:
\begin{itemize}
	\item an $\ell_{\infty}$-ball with radius $\bm{\epsilon$} and centroid $\xnom$, s.t. $\Xtt=\binfxnom$; or
	\item a polytope for a specified $\Aout\in\mathds{R}^{\mout\times n_x}$, meaning we will compute $\mathbf{b}^\text{out}\in\mathds{R}^{m_{out}}$ s.t. $\Xtt=\{\mathbf{x}_t \in \mathds{R}^{n_x}\ \lvert\ \Aout\mathbf{x}_t\leq \mathbf{b}^\text{out}\}$.
\end{itemize}

We assume that either $\Aout$ is provided (in the case of polytope output bounds), or that $\Aout=\mathbf{I}_{n_x}$ (in the case of $\ell_{\infty}$ output bounds).
Note that we use $\ifacet$ to index polytope facets, $\icontrol$ to index the control vectors, and $\istate$ to index the state vectors.
In this section, we assume that $\mathcal{U}_t=\mathds{R}^{n_u}$ (no control input constraints) for cleaner notation; this assumption is relaxed in~\cref{sec:forward_reachability:control_limits}.

The exact 1-step closed-loop reachability problem is as follows. For each row $\ifacet$ in $\Aout$, solve the following optimization problem,
\begin{align}
\begin{split}
    \left(\boutj\right)^* = \max_{\mathbf{x}_t\in\Xt} &\quad \Aoutj \mathbf{x}_{t+1} \\
    \text{s.t.} &\quad \mathbf{x}_{t+1} = f(\mathbf{x}_t; \pi), \label{eqn:nfl_analysis_exact}
\end{split}
\end{align}
where $\left(\bout\right)^*$ defines the tightest description of $\Xtt$ associated with $\Aout$.
However, the nonlinearities in $\pi$ make solving this problem intractable in practice, so this section describes a relaxation that provides bounds on~\cref{eqn:nfl_analysis_exact}.

\subsection[Bounds on Next State from a particular state]{Bounds on $\mathbf{x}_{t+1}$ from a particular $\mathbf{x}_t$}\label{sec:forward_reachability:particular_xt}

% The $B_t$ matrix projects the set of controls $\Pi=[\pi^L_1(\y_t), \pi^U_1(\y_t)]\times\ldots\times[\pi^L_{n_u}(\y_t), \pi^U_{n_u}(\y_t)]$ into the space of next states.

\begin{lemma}\label[lemma]{thm:particular_xt}
Given an $m$-layer NN control policy $\pi:\R^{n_y}\to\R^{n_u}$, closed-loop dynamics $f: \R^{n_x} \times \Pi \to \R^{n_x}$ as in~\cref{eqn:ltv_dynamics,eqn:closed_loop_dynamics}, and specification matrix $\Aout\in\R^{\mout\times n_x}$, for each $\ifacet\in [\mout]$, there exist two explicit functions $\piLCL: \R^{n_y}\to\R^{n_u}$ and $\piUCL: \R^{n_y}\to\R^{n_u}$ such that $\forall \icontrol\in [n_m], \forall \x_t\in\mathcal{B}_p(\x_{t,0}, \bm{\epsilon)}$ and $\forall \y_t\in\mathcal{B}_\infty(\mathbf{C}_t^T\x_t + \frac{\bar{\bm{\nu}}_t+\ubar{\bm{\nu}}_t}{2}, \frac{\bar{\bm{\nu}}_t-\ubar{\bm{\nu}}_t}{2})$, the inequality $\Aoutj f(\x_t, \piLCL) \leq \Aoutj f(\x_t, \pi) \leq \Aoutj f(\x_t, \piUCL)$ holds true, where
\begin{align}
    \piUCL(\y_t) &= \piclAu_{\ifacet,:,:}\y_t+ \piclbu_{:,\ifacet} \label{eqn:piucl}\\
    \piLCL(\y_t) &= \piclAl_{\ifacet,:,:}\y_t+ \piclbl_{:,\ifacet} \label{eqn:pilcl},
\end{align}
letting $\piclAu,\piclAl \in\R^{\mout\times n_u \times n_y}$ and $\piclbu, \piclbl \in\R^{n_u \times \mout}$,
\begin{align}
    \piclAu_{\ifacet,:,:} &= \Jbar{\AoutjBt}{\CROWNAu}{\CROWNAl} \\
    \piclAl_{\ifacet,:,:} &= \Jbar{\AoutjBt}{\CROWNAl}{\CROWNAu} \\
    \piclbu_{:,\ifacet} &= \Jbar{\AoutjBt}{\CROWNbu}{\CROWNbl} \\
    \piclbl_{:,\ifacet} &= \Jbar{\AoutjBt}{\CROWNbl}{\CROWNbu},
\end{align}
using selector function (similar to \texttt{torch.where}) $s: \R^m \times \R^{m\times n} \times \R^{m \times n} \to \R^{m \times n}$, where the matrix returned by $s$ is defined element-wise $\forall a \in [m], b \in [n]$,
\begin{align}
\left[\Jbar{\mathbf{z}}{\mathbf{A}}{\mathbf{B}}\right]_{a,b} &=
\begin{cases}
\mathbf{A}_{a,b}, & \textrm{if } \mathbf{z}_{a} \geq 0 \\
\mathbf{B}_{a,b}, & \textrm{otherwise}
\end{cases}
\end{align}
and where $\CROWNAu, \CROWNAl, \CROWNbu, \CROWNbl$ are computed from~\cref{thm:crown_particular_x} with $\y_0=\mathbf{C}_t^T(\x_{t,0}+\frac{\bar{\bm{\nu}}_t+\ubar{\bm{\nu}}_t}{2})$, and $\bm{\epsilon=}\bm{\epsilon+}\frac{\bar{\bm{\nu}}_t-\ubar{\bm{\nu}}_t}{2}$.

\begin{proof}
For any particular measurement $\y_t$, after relaxing the NN according to~\cref{thm:crown_particular_x}, let $\Pi(\y_t)=\{ \pi \lvert \pi_\icontrol^L(\y_t) \leq \pi_\icontrol (\y_t) \leq \pi_\icontrol ^U(\y_t) \forall \icontrol \in[n_u] \}$ denote the set of possible effective control policies.
Denote the control policy $\piUCL\in\Pi(\y_t)$ as one that induces the least upper bound on the $\ifacet$-th facet of the next state polytope,
\begin{align}
    \Aoutj f(\mathbf{x}_{t}; \pi_{:,\ifacet}^{U_{CL}}) =& \max_{\pi \in \Pi(\y_t)} \Aoutj  f(\mathbf{x}_t; \pi) \notag \\
    & \hspace*{-1.1in} = \max_{\pi \in \Pi(\y_t)} \Aoutj \left[\mathbf{A}_{t} \mathbf{x}_t +  %\right. \notag\\
    %&\quad\quad\quad\quad\quad\quad\left. 
    \mathbf{B}_{t} \pi(\y_t) + \mathbf{c}_{t} + \bm{\omega}_t\right] \nonumber\\
    &\hspace*{-1.1in} = \left[\max_{\pi \in \Pi(\y_t)} \Aoutj \mathbf{B}_{t} \pi(\y_t) \right] + %\\
    %&\quad\quad 
    \Aoutj \left[\mathbf{A}_{t} \mathbf{x}_t + \mathbf{c}_{t} + \bm{\omega}_t\right],% \nonumber
\end{align}
Thus for $\y_t$,
\begin{align}
    \pi_{:,\ifacet}^{U_{CL}} =& \argmax_{\pi \in \Pi(\y_t)} \Aoutj  \mathbf{B}_{t} \pi(\y_t).
\end{align}
The resulting control input $\forall \ifacet\in[m_{t+1}],\icontrol\in[n_u]$ is,
\begin{align}
    \pi_{\icontrol,\ifacet}^{U_{CL}}(\y_t) =& \begin{cases}
        \pi_\icontrol^U(\y_t), & \mathrm{if}\ \Aoutj  \mathbf{B}_{t,:,\icontrol} \geq 0 \\
        \pi_\icontrol^L(\y_t), & \mathrm{otherwise}
    \end{cases}
    \label{eqn:pi_upoly_elementwise}.
\end{align}
Writing \cref{eqn:pi_upoly_elementwise} in matrix form results in~\cref{eqn:piucl}.
The proof of the lower bound follows similarly.
\end{proof}
\end{lemma}

\subsection[Bounds on Next state from initial state set]{Bounds on $\x_{t+1}$ from any $\xinX$}\label{sec:forward_reachability:bounds_xt1}

Now that we can bound each facet of the next state polytope given a particular current state and observation, we can form bounds on the next state polytope facet given a \textit{set} of possible current states.
This is necessary to handle initial state set constraints and to compute ``$t>1$''-step reachable sets recursively as in~\cref{eqn:reachable_sets}.
We assume $\xinX$.

\begin{lemma}\label[lemma]{thm:bounds_on_xt1}
Given an $m$-layer NN control policy $\pi:\R^{n_y}\to\R^{n_u}$, closed-loop dynamics $f: \R^{n_x} \times \Pi \to \R^{n_x}$ as in~\cref{eqn:ltv_dynamics,eqn:closed_loop_dynamics}, and specification matrix $\Aout\in\R^{\mout\times n_x}$, for each $\ifacet\in[\mout]$, there exist two fixed values $\xttubifacet $ and $\xttlbifacet$ such that $\forall \xinX$, the inequality $\xttlbifacet \leq \Aoutj f(\x_t; \pi) \leq \xttubifacet $ holds true, where
\begin{align}
    \xttubifacet &= \max_{\xinX} \Muj \mathbf{x}_t + \Nuj \label{eqn:global_upper_bnd_generic_optimization} \\
    \xttlbifacet &= \min_{\xinX} \Mlj \mathbf{x}_t + \Nlj\label{eqn:global_lower_bnd_generic_optimization},
\end{align}
with $\mathbf{M}^U\in\R^{n_x \times n_x}$, $\mathbf{n}^U\in\R^{n_x}$ defined as
\begin{align}
    \Muj &= \left(\Aoutj  \left(\mathbf{A}_{t} + \mathbf{B}_{t} \piclAu_{\ifacet,:,:} \mathbf{C}_t^T \right) \right) \label{eqn:muj_defn}\\
    \Mlj &= \left(\Aoutj  \left(\mathbf{A}_{t} + \mathbf{B}_{t} \piclAl_{\ifacet,:,:} \mathbf{C}_t^T \right) \right) \\
    \Nuj &= \Aoutj \left(\mathbf{B}_{t}\left( \piclAu_{\ifacet,:,:} \Jbar{\AoutjBtPiclAuCt}{\bar{\bm{\nu}}_t}{\ubar{\bm{\nu}}_t} + \piclbu_{:,\ifacet} \right) + \nonumber\right.\\ &\quad\quad\quad\quad\left.\mathbf{c}_{t} + \Jbar{\AoutjBt}{\bar{\bm{\omega}}_t}{\ubar{\bm{\omega}}_t} \right) \label{eqn:nuj_defn}\\
    \Nlj &= \Aoutj \left(\mathbf{B}_{t}\left( \piclAl_{\ifacet,:,:} \Jbar{\AoutjBtPiclAlCt}{\ubar{\bm{\nu}}_t}{\bar{\bm{\nu}}_t} + \piclbl_{:,\ifacet} \right) + \nonumber\right.\\ &\quad\quad\quad\quad\left.\mathbf{c}_{t} + \Jbar{\AoutjBt}{\ubar{\bm{\omega}}_t}{\bar{\bm{\omega}}_t} \right),
\end{align}
with $\piclAu, \piclAl, \piclbu, \piclbl$ computed from~\cref{thm:particular_xt}.

\begin{proof}
Bound the next state polytope's $\ifacet$-th facet above,
\begingroup
\allowdisplaybreaks
\begin{align}
    \Aoutj &\mathbf{x}_{t+1}=\Aoutj f(\mathbf{x}_{t};\pi) \\
    &\leq \Aoutj f(\mathbf{x}_{t}; \piUCL) \\
    &\leq \max_{\xinX}\Aoutj f(\mathbf{x}_{t}; \piUCL) := \xttubifacet \\
    &= \max_{\xinX} \Aoutj  \left[\mathbf{A}_{t} \mathbf{x}_t + \mathbf{B}_{t} \piUCL(\y_t)+\mathbf{c}_{t} + \bm{\omega}_t \right]\\
    &= \max_{\xinX} \Aoutj \left[\mathbf{A}_{t} \mathbf{x}_t + \mathbf{B}_{t} \left(\piclAu_{\ifacet,:,:}\y_t+\piclbu_{:,\ifacet}\right) + \nonumber\right.\\&\quad\quad\quad\quad\quad\quad\quad\left.\mathbf{c}_{t} + \bm{\omega}_t \right] \label{eqn:global_bnd_sub_piUCL} \\ 
    &= \max_{\xinX} \Aoutj \left[\mathbf{A}_{t} \mathbf{x}_t + \mathbf{B}_{t} \left(\piclAu_{\ifacet,:,:}\left(\mathbf{C}_t^T\x_t+\bm{\nu}_t\right)+\piclbu_{:,\ifacet}\right) \nonumber\right.\\&\quad\quad\quad\quad\quad\quad\quad\left.+ \mathbf{c}_{t} + \bm{\omega}_t \right] \label{eqn:global_bnd_observation} \\
    &= \max_{\xinX} \left(\Aoutj \left(\mathbf{A}_{t} + \mathbf{B}_{t} \piclAu_{\ifacet,:,:} \mathbf{C}_t^T\right) \right) \x_t + \nonumber\\& \quad\quad\quad\Aoutj\left( \mathbf{B}_{t} \left(\piclAu_{\ifacet,:,:} \bm{\nu}_t + \piclbu_{:,\ifacet} \right) + \mathbf{c}_{t} + \bm{\omega}_t \right) \label{eqn:global_bnd_linear_in_x} \\
    &= \max_{\xinX} \left(\Aoutj \left(\mathbf{A}_{t} + \mathbf{B}_{t} \piclAu_{\ifacet,:,:} \mathbf{C}_t^T\right) \right) \x_t + \nonumber\\& \quad\quad\quad\Aoutj\left( \mathbf{B}_{t} \left(\piclAu_{\ifacet,:,:} \Jbar{\AoutjBtPiclAuCt}{\bar{\bm{\nu}}_t}{\ubar{\bm{\nu}}_t} + \piclbu_{:,\ifacet} \right) + \right.\nonumber\\&\quad\quad\quad\quad\left. \mathbf{c}_{t} + \Jbar{\AoutjBt}{\bar{\bm{\omega}}_t}{\ubar{\bm{\omega}}_t} \right), \label{eqn:global_bnd_worst_noise}
\end{align}
\endgroup
where \cref{eqn:global_bnd_sub_piUCL} substitutes the definition of $\piUCL$ from~\cref{thm:particular_xt}, \cref{eqn:global_bnd_observation} substitutes the observation from~\cref{eqn:closed_loop_dynamics}, \cref{eqn:global_bnd_linear_in_x} separates terms that depend on $\x_t$, and \cref{eqn:global_bnd_worst_noise} introduces the worst-case realizations of process and measurement noise.
Substituting $\Muj, \Nuj$ results in~\cref{eqn:global_upper_bnd_generic_optimization}.
The proof of the lower bound follows similarly.
\end{proof}

\end{lemma}

% \subsection[Solving Optimization Problems]{Solving~\cref{eqn:global_upper_bnd_generic_optimization,eqn:global_lower_bnd_generic_optimization}}\label{sec:forward_reachability:solving_opt_problems}

The optimization problems in~\cref{eqn:global_upper_bnd_generic_optimization,eqn:global_lower_bnd_generic_optimization} have convex cost with convex constraints $\xinX$ (e.g., polytope $\Xt$).
% \begin{align}
%     \max_{\xinX} \Aoutj  f(\mathbf{x}_t; \piUCL) &= \underbrace{\left(\max_{\Ainbin} \Muj\mathbf{x}_t \right)}_{\mathbf{m}_{\ifacet}^{U}\leftarrow \texttt{LP\_SOLVER}} + \Nuj\\
%     &= \mathbf{m}_{\ifacet}^{U} + \Nuj,
% \end{align}
We solve the linear programs (LPs) with \texttt{cvxpy}~\cite{diamond2016cvxpy},
\begin{align}
    \xttubifacet  &= \texttt{LP}(\Muj\mathbf{x}_t, \Ain, \mathbf{b}^\text{in})+\Nuj \label{eqn:lp_upper_bnd}\\
    \xttlbifacet &= \texttt{LP}(-\Mlj\mathbf{x}_t, \Ain, \mathbf{b}^\text{in})+\Nlj \label{eqn:lp_lower_bnd}.
\end{align}

\subsection{Converting State Constraints into Reachable Sets}\label{sec:forward_reachability:reachable_sets}

\subsubsection[Reachable Sets as l-inf balls]{Reachable Sets as $\ell_{\infty}$-balls}

Assume $\mathcal{X}_0$ is an $\ell_p$-ball.
Define $\{p,\bm{\epsilon,}\xnom\}$ s.t. $\mathcal{X}_0\subseteq \bpxnom$ and let $\bar{\mathcal{R}}_0(\mathcal{X}_0)=\mathcal{X}_0$.
Using the results of the previous section, use $\mathbf{x}_{t=0} \in\bpxnom$ to compute $(\xttubonelp,\, \xttlbonelp)$ for each index of the state vector $\istate\in[n_x]$, specifying $\Aout=\mathbf{I}_{n_x}$.
Recursively compute
\begin{align} \hspace*{-1.5mm}
    \bar{\mathcal{R}}_{t+1}(\mathcal{X}_0)=\mathcal{B}_{\infty}\left(\frac{\xttub + \xttlb}{2}, \frac{\xttub - \xttlb}{2}\right) \label{eqn:reachable_set_linf}.
\end{align}

\subsubsection{Reachable Sets as Polytopes}

Assume $\mathcal{X}_0$ is an $\ell_p$-ball or polytope.
Either define $\{p,\bm{\epsilon,}\xnom\}$ s.t. $\mathcal{X}_0\subseteq\bpxnom$ or define $\{\Ain, \mathbf{b}^\text{in}\}$ s.t. $\Xt=\{\mathbf{x}_t \lvert \Ainbin\}$ and let $\bar{\mathcal{R}}_0(\mathcal{X}_0)=\mathcal{X}_0$.
Using the results of the previous section, use $\mathbf{x}_{t=0} \in\bpxnom$ or 
$\{\Ain, \mathbf{b}^\text{in}\}$ to compute $(\xttubonepoly,\, \xttlbonepoly)$ for each index of output polytope facets $\ifacet\in[m_{out}]$, giving
\begin{align}
    \bar{\mathcal{R}}_{t+1}(\mathcal{X}_0) = \left\{ \mathbf{x}_{t}\ \lvert\ \begin{bmatrix} \Aout \\ -\Aout \end{bmatrix} \mathbf{x}_t \leq \begin{bmatrix} \xttub \\ -\xttlb \end{bmatrix} \right\} \label{eqn:reachable_set_polytope}.
\end{align}
In both cases, $\bar{\mathcal{R}}_{t}(\mathcal{X}_0)\supseteq\mathcal{R}_{t}(\mathcal{X}_0) \forall t\geq0$, so these $\bar{\mathcal{R}}_{t}$ can be used to verify the original closed loop system~\cref{eqn:closed_loop_dynamics}.

\subsection{Closed-Form Solution}\label{sec:forward_reachability:closed_form_soln}

Rather than employing an LP solver as in~\cref{eqn:lp_upper_bnd}, the optimization problem in~\cref{eqn:global_upper_bnd_generic_optimization} can be solved in closed-form when $\Xt$ is described by an $\ell_p$-ball.

\begin{lemma}\label[lemma]{thm:closed_form}
In the special case of~\cref{thm:bounds_on_xt1} where $\Xt = \bpxtnom$ for some $p\in[1,\infty)$, the following closed-form expressions are equivalent to~\cref{eqn:global_upper_bnd_generic_optimization,eqn:global_lower_bnd_generic_optimization}, respectively,
\begin{align}
    \xttubistate &= \lvert\lvert\bm{\epsilon}\odot \Muj \rvert\rvert_q + \Muj \xtnom + \Nuj \label{eqn:closed_form_upper_bnd}\\
    \xttlbistate &= -\lvert\lvert\bm{\epsilon}\odot \Mlj \rvert\rvert_q + \Mlj \xtnom + \Nlj \label{eqn:closed_form_lower_bnd},
\end{align}
where ${1/p + 1/q = 1}$ (e.g., $p=\infty, q=1$).

\begin{proof}
The proof proceeds as in \cite{everett2020certified} and \cite{Weng_2018}.
\begin{align}
\xttubistate &= \max_{\xinXball} \Muj \mathbf{x}_t + \Nuj \label{eq:qlj_s} \\
&= \left(\max_{\bm{\zeta}\in \mathcal{B}_p(\bm{0}, \bm{1})}\Muj (\bm{\zeta}\odot\bm{\epsilon})\right) + \Muj \xtnom + \Nuj \label{eq:qlj_y} \\
&= \left(\max_{\bm{\zeta}\in \mathcal{B}_p(\bm{0}, \bm{1})}(\bm{\epsilon}\odot \Muj) \bm{\zeta}\right) + \Muj \xtnom + \Nuj \label{eq:qlj_shuffle} \\
&= \lvert\lvert\bm{\epsilon}\odot \Muj \rvert\rvert_q + \Muj \xtnom + \Nuj \label{eq:qlj_norm},
\end{align}
with $\odot$ denoting element-wise multiplication (Hadamard product).
Recall that $\Nuj$ does not depend on $\mathbf{x}_t$.
From~\cref{eq:qlj_s} to~\cref{eq:qlj_y}, we substitute ${\mathbf{x}_t := \bm{\zeta} \odot \bm{\epsilon} + \xtnom}$, to shift and re-scale the observation to within the unit ball around zero, $\bm{\zeta}\in \mathcal{B}_p(\bm{0},\bm{1})$.
The commutative property of the Hadamard product allows \cref{eq:qlj_y} to \cref{eq:qlj_shuffle}.
The maximization in~\cref{eq:qlj_shuffle} is equivalent to a $\ell_q$-norm in~\cref{eq:qlj_norm} by the definition of the dual norm $\lvert\lvert \bm{a} \rvert\rvert_q = \{\text{sup}\ \bm{a}^T \bm{b}:\lvert\lvert \bm{b} \rvert\rvert_p \leq 1\}$ and the fact that the $\ell_q$ norm is the dual of the $\ell_p$ norm for $p,q \in [1,\infty)$ (with ${1/p + 1/q = 1}$). The proof of the lower bound follows similarly.
\end{proof}
\end{lemma}

% \subsection{n-step improvement}

% \mfe{Show how $n$-step process works in some cases (new lemma?)}
% \mfe{Add a plot with n-step improvement.}

\subsection{Algorithm for Computing Forward Reachable Sets}

\begin{algorithm}
    \caption{Closed-Loop CROWN Propagator}
    \begin{algorithmic}[1]
    \setcounter{ALC@unique}{0}
    \renewcommand{\algorithmicrequire}{\textbf{Input:}}
    \renewcommand{\algorithmicensure}{\textbf{Output:}}
    \REQUIRE initial state set $\mathcal{X}_0$, trained NN control policy $\pi$, dynamics $f$, time horizon $\tau$
    \ENSURE forward reachable set approximations $\bar{\mathcal{R}}_{1:\tau}(\mathcal{X}_0)$
    \STATE $\bar{\mathcal{R}}_0(\mathcal{X}_0) \gets \mathcal{X}_0$ \label{alg:cl_crown_propagator:initialize}
    \FOR{$t \in \{ 1, 2, \ldots, \tau \}$} \label{alg:cl_crown_propagator:timestep_for_loop}
        \STATE $\mathcal{Y}_{t-1} \gets \text{possibleObs}(\bar{\mathcal{R}}_{t-1}(\mathcal{X}_0), f)$
        \STATE $\CROWNAu, \CROWNAl, \CROWNbu, \CROWNbl \leftarrow \mathrm{CROWN}(\pi, \mathcal{Y}_{t-1})$
        \FOR{$\ifacet \in [\mout]$ or $\istate \in [n_x]$}
            \STATE $\xttubifacet \leftarrow$ \cref{eqn:lp_upper_bnd} or \cref{eqn:closed_form_upper_bnd}
            \STATE $\xttlbifacet \leftarrow$ \cref{eqn:lp_lower_bnd} or \cref{eqn:closed_form_lower_bnd}
        \ENDFOR
        \\ \texttt{// Polytope Reachable Sets}
        \STATE $\bar{\mathcal{R}}_t(\mathcal{X}_0) \leftarrow \left\{ \mathbf{x}\ \lvert\ \begin{bmatrix} \Aout \\ -\Aout \end{bmatrix} \mathbf{x} \leq \begin{bmatrix} \xttub \\ -\xttlb \end{bmatrix} \right\}$ from \cref{eqn:reachable_set_polytope}
        \\ \texttt{// $\ell_\infty$-ball Reachable Sets}
        \STATE $\bar{\mathcal{R}}_t(\mathcal{X}_0) \leftarrow \mathcal{B}_{\infty}\left(\frac{\xttub + \xttlb}{2}, \frac{\xttub - \xttlb}{2}\right)$ from \cref{eqn:reachable_set_linf}
    \ENDFOR
    \RETURN $\bar{\mathcal{R}}_{1:\tau}(\mathcal{X}_0)$ \label{alg:cl_crown_propagator:return}
\end{algorithmic}\label{alg:cl_crown_propagator}
\end{algorithm}

The proposed procedure for estimating forward reachable sets from $\mathcal{X}_0$ for a neural feedback loop is provided in~\cref{alg:cl_crown_propagator} (the term ``propagator'' will be explained in~\cref{sec:partition}).
After initializing the zeroth forward reachable set as the initial set (\cref{alg:cl_crown_propagator:initialize}), we recursively compute $\bar{\mathcal{R}}_{t}(\mathcal{X}_0)$ forward in time (\cref{alg:cl_crown_propagator:timestep_for_loop}).
For each timestep, we determine which observations the NN could receive, then pass those through CROWN~\cite{zhang2018efficient} (or another NN relaxation method, provided it returns the corresponding terms).
Depending on whether polytope facets or $\ell_p$-balls are used, the upper/lower bounds are computed facet-by-facet or state-by-state.
Then, the $\bm{\gamma}$ terms are converted into set representations as in~\cref{sec:forward_reachability:reachable_sets}.
After following this procedure for each timestep, the algorithm returns a sequence of outer-approximations of the forward reachable sets (\cref{alg:cl_crown_propagator:return}).

\section{Tighter Reachable Sets by Partitioning the Initial State Set}\label{sec:partition}

NN relaxation methods can be improved by partitioning the input set~\cite{xiang2018output}, particularly when the input set is large but of low dimension.
Here, we achieve tighter bounds by splitting $\mathcal{X}_0$ into subsets, computing $N$-step reachable sets for each of the subsets separately, then returning the union of all reachable sets from each subset.

Recall that~\cite{everett2020robustness} introduced an architecture composed of a \textit{partitioner} and \textit{propagator} for analyzing NNs in isolation.
The partitioner is an algorithm to split the NN input set in an intelligent way (e.g., uniform gridding~\cite{xiang2018output}, simulation guidance~\cite{xiang2020reachable}, greedy simulation guidance~\cite{everett2020robustness}), and the propagator is an algorithm to estimate bounds on the NN outputs given a NN input set (e.g., IBP~\cite{gowal2018effectiveness}, Fast-Lin~\cite{Weng_2018}, CROWN~\cite{zhang2018efficient}, SDP~\cite{fazlyab2019safety}).
The partitioner can query the propagator repeatedly to refine the estimated output set bounds.
The choice and design of partitioners and propagators has important implications on the bound tightness and computational runtime.

\begin{figure}[t]
    \includegraphics[page=5,width=\linewidth,trim=10 120 0 70,clip]{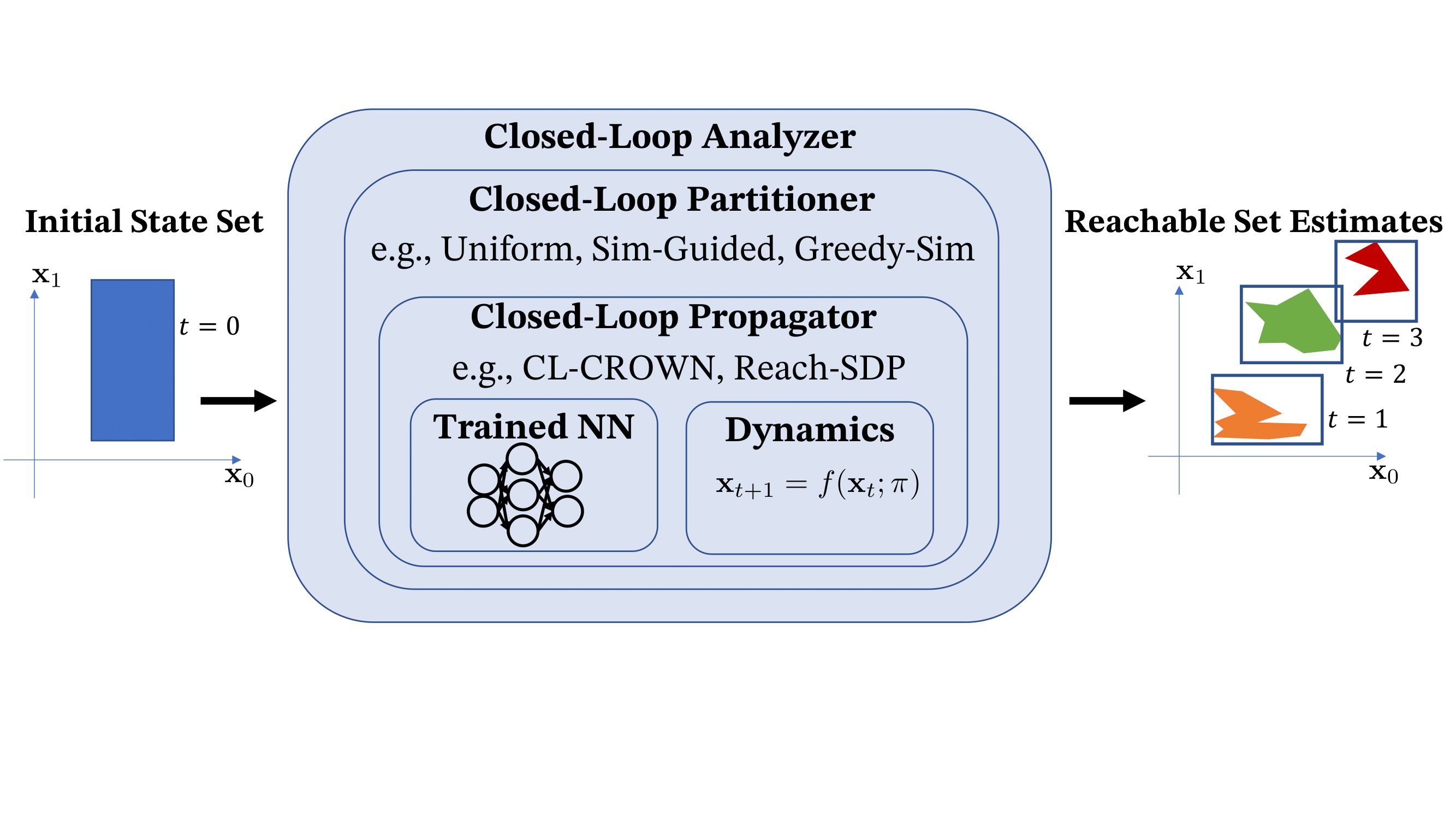}
    \caption{System Architecture. A closed-loop propagator uses the trained NN control policy and dynamics to estimate reachable sets, and a closed-loop partitioner decides how to split the initial state set into pieces. This is the closed-loop extension of the architecture from our prior work~\cite{everett2020robustness}.}
    \label{fig:system_architecture}
\end{figure}

This work extends the framework from~\cite{everett2020robustness} to closed-loop systems, as visualized in~\cref{fig:system_architecture}.
In particular, CL-CROWN and Reach-SDP represent \textit{closed-loop propagators} (given an initial state set, they compute reachable sets for a trained NN control policy and closed-loop dynamics), and in this section we extend partitioners discussed in~\cite{everett2020robustness} to be \textit{closed-loop partitioners}.
Altogether, we call the nested architecture a \textit{closed-loop analyzer}.

\subsection{Closed-Loop Partitioners}

The simplest closed-loop partitioner splits the input set into a uniform grid, as described in~\cref{alg:cl_uniform_partitioner}.
After splitting $\mathcal{X}_0$ into $\Pi_{\istate=0}^{n_x}\mathbf{r}_\istate$ cells (\cref{alg:cl_uniform_partitioner:partition}), each cell of $\mathcal{X}_0$ is passed through a closed-loop propagator, CLProp (e.g., CL-CROWN (\cref{alg:cl_crown_propagator}) or Reach-SDP~\cite{hu2020reach}).
For each timestep, the returned estimate of the reachable set for all of $\mathcal{X}_0$ is thus the union of all reachable sets for cells of the $\mathcal{X}_0$ partition.

\begin{algorithm}[t]
    \caption{Closed-Loop Uniform Partitioner}
    \begin{algorithmic}[1]
    \setcounter{ALC@unique}{0}
    \renewcommand{\algorithmicrequire}{\textbf{Input:}}
    \renewcommand{\algorithmicensure}{\textbf{Output:}}
    \REQUIRE initial state set $\mathcal{X}_0$, trained NN control policy $\pi$, dynamics $f$, time horizon $\tau$, number of partition cells $\mathbf{r}$
    \ENSURE forward reachable set approximations $\bar{\mathcal{R}}_{1:\tau}(\mathcal{X}_0)$
    \STATE $\bar{\mathcal{R}}_{t} \gets \emptyset \quad \forall t \in \{1,\ldots,\tau\}$
    \STATE $\mathcal{S} \gets \text{partition}(\mathcal{X}_0, \mathbf{r})$ \label{alg:cl_uniform_partitioner:partition}
    \FOR{$\mathcal{X}' \in \mathcal{S}$}
        \STATE $\bar{\mathcal{R}}_{1:\tau}' \gets \text{CLProp}(\mathcal{X}', f, \pi, \tau)$
        \STATE $\bar{\mathcal{R}}_t \gets \bar{\mathcal{R}}_t \cup \bar{\mathcal{R}}_{t}' \quad \forall t \in \{1,\ldots,\tau\}$
    \ENDFOR
    \RETURN $\bar{\mathcal{R}}_{1:\tau}(\mathcal{X}_0)$
\end{algorithmic}\label{alg:cl_uniform_partitioner}
\end{algorithm}

There are several reasons to use an iterative partitioner rather than the uniform partitioner presented before, as discussed in~\cite{xiang2020reachable,everett2020robustness}.
Thus, the closed-loop variant of the greedy simulation-guided partitioner from~\cite{everett2020robustness} is described in~\cref{alg:cl_gsg_partitioner}.
After acquiring $N$ Monte Carlo samples from $\mathcal{X}_0$ (\cref{alg:cl_gsg_partitioner:mc}), those sampled points are simulated forward in time according to the dynamics and control policy (\cref{alg:cl_gsg_partitioner:mc_dyn}).
The extrema of these samples can be used to define the sampled reachable sets at each timestep (\cref{alg:cl_gsg_partitioner:mc_reach}).
Then, the full initial state set $\mathcal{X}_0$ is passed through a closed-loop propagator (\cref{alg:cl_gsg_partitioner:clprop_full}) and the input and reachable set estimates are added to a stack, $M$.

An iterative process is used to refine the reachable set estimates from the full $\mathcal{X}_0$.
The element with largest distance from the sampled reachable set estimate is popped from $M$ (e.g., use the same idea from Fig. 3a of~\cite{everett2020robustness} on the final timestep's reachable set).
If the chosen element's reachable set is within the sampled reachable set for all timesteps, there is no need to further refine that cell and its reachable set can be added to set that will eventually be returned.
Otherwise, the input set of that element is bisected (\cref{alg:cl_gsg_partitioner:bisect}) and each of the two input sets are passed through CLProp and added to $M$.
Termination conditions, such as an empty $M$ or number of CLProp calls or a timeout can be implemented depending on the application.
Finally, the remaining elements of the stack are added to the set that will be returned.

\begin{algorithm}[t]
    \caption{Closed-Loop Greedy Sim-Guided Partitioner}
    \begin{algorithmic}[1]
    \setcounter{ALC@unique}{0}
    \renewcommand{\algorithmicrequire}{\textbf{Input:}}
    \renewcommand{\algorithmicensure}{\textbf{Output:}}
    \REQUIRE initial state set $\mathcal{X}_0$, number of MC samples $N$, trained NN control policy $\pi$, dynamics $f$, time horizon $\tau$
    \ENSURE forward reachable set approximations $\bar{\mathcal{R}}_{1:\tau}(\mathcal{X}_0)$
    \STATE $\mathcal{X}_0^\text{MC} \stackrel{N\ \text{i.i.d.}}{\sim} \text{Unif}(\mathcal{X}_0)$ \label{alg:cl_gsg_partitioner:mc}
    \STATE $\mathcal{X}_{t+1}^{MC} \gets f(\mathcal{X}_t^\text{MC}; \pi) \quad \forall t\in\{1,\ldots,\tau\}$ \label{alg:cl_gsg_partitioner:mc_dyn}
    \STATE $\mathcal{R}_t^\text{MC} \gets \text{extrema of } \mathcal{X}_t^\text{MC} \quad \forall t \in \{1,\ldots,\tau\}$ \label{alg:cl_gsg_partitioner:mc_reach}
    \STATE $\bar{\mathcal{R}}_{1:\tau} \gets \text{CLProp}(\mathcal{X}_0, f, \pi, \tau)$ \label{alg:cl_gsg_partitioner:clprop_full}
    \STATE $M \gets \{ (\mathcal{X}_0, \bar{\mathcal{R}}_{1:\tau}) \}$
    \WHILE{$M\neq \emptyset$ or termination condition}
        \STATE $(\mathcal{X}', \mathcal{R}'_{1:\tau}) \stackrel{\text{M.pop}}{\gets} \argmax_{(\mathcal{X}'', \mathcal{R}''_{1:\tau})\in M}{d(\mathcal{R}^\text{MC}_{1:\tau}, \mathcal{R}''_{1:\tau})}$
        \IF{$\mathcal{R}'_{t} \subseteq \mathcal{R}^\text{MC}_{t} \quad \forall t\in\{1,\ldots,\tau\}$}
            \STATE $\bar{\mathcal{R}}_t \gets \bar{\mathcal{R}}_t \cup \mathcal{R}'_t \quad \forall t\in\{1,\ldots,\tau\}$
        \ELSE
            \STATE $\mathcal{X}^1, \mathcal{X}^2 \gets \text{Bisect}(\mathcal{X}')$ \label{alg:cl_gsg_partitioner:bisect}
            \STATE $\bar{\mathcal{R}}_{1:\tau}^1 \gets \text{CLProp}(\mathcal{X}^1, f, \pi, \tau)$
            \STATE $\bar{\mathcal{R}}_{1:\tau}^2 \gets \text{CLProp}(\mathcal{X}^2, f, \pi, \tau)$
            \STATE $M \gets M \cup \{ (\mathcal{X}^1, \bar{\mathcal{R}}_{1:\tau}^1), (\mathcal{X}^2, \bar{\mathcal{R}}_{1:\tau}^2) \}$
        \ENDIF
    \ENDWHILE
    \STATE $\bar{\mathcal{R}}_t \gets \bar{\mathcal{R}}_t \cup (\cup_{(\mathcal{X}'', \mathcal{R}'')\in M} \mathcal{R}''_t) \quad \forall t \in \{1,\ldots,\tau\}$
    \RETURN $\bar{\mathcal{R}}_{1:\tau}(\mathcal{X}_0)$
\end{algorithmic}\label{alg:cl_gsg_partitioner}
\end{algorithm}

\section{Handling Nonlinearities}\label{sec:nonlinear}

\subsection[Accounting for Control Limits]{Accounting for Control Limits, $\mathcal{U}_t$}\label{sec:forward_reachability:control_limits}

\begin{figure}[t]
    \centering
    \includegraphics[page=3,width=\linewidth,trim=0 0 270 100,clip]{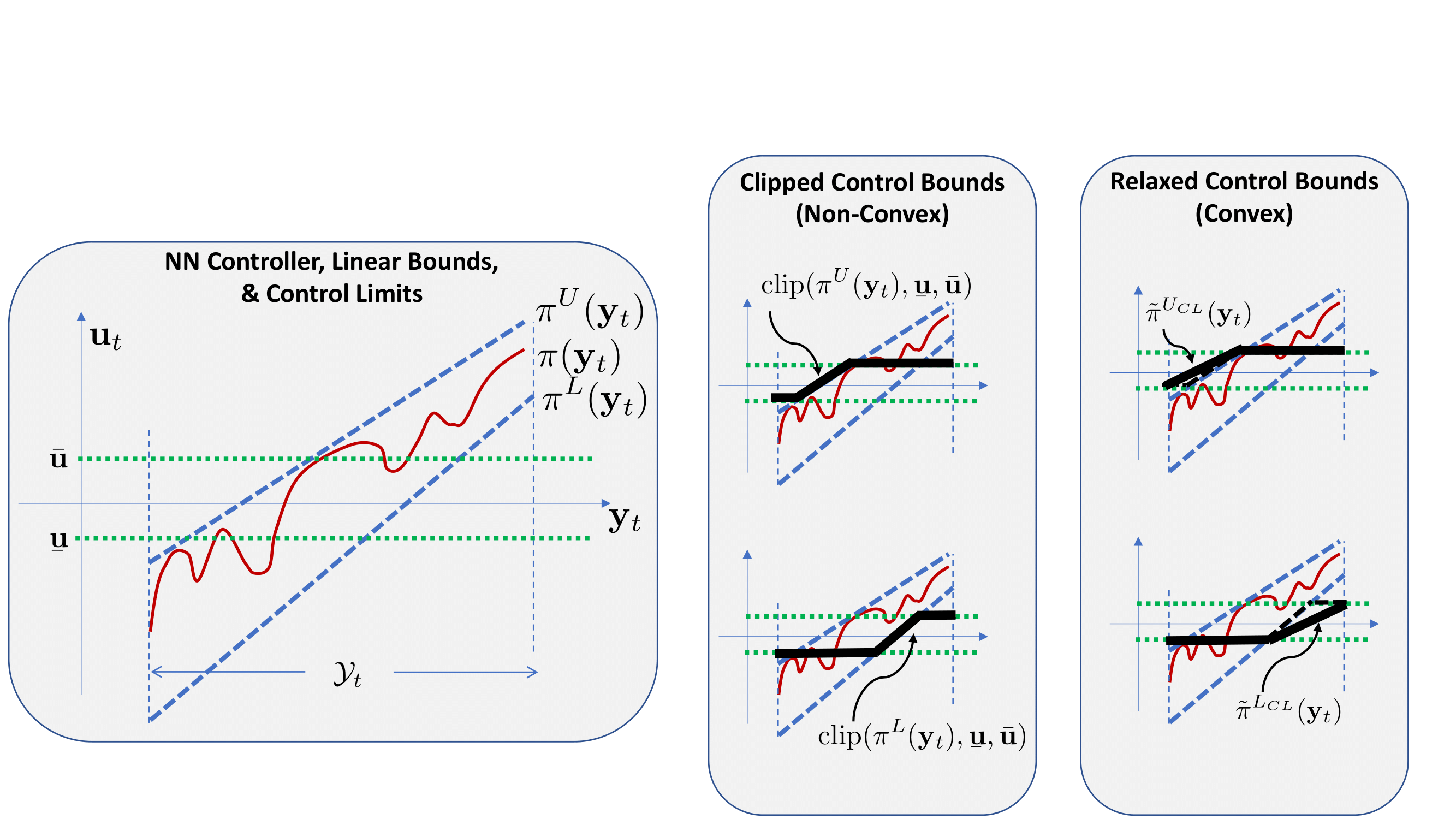}
    \caption{Control Limits. The NN control policy (red) is bounded by $\pi^U, \pi^L$ from CROWN (blue), but none of these respect the control limits (green). While adding a $\mathrm{clip}$ could be non-convex (right block), \cref{sec:forward_reachability:control_limits} borrows the idea from~\cite{hu2020reach} to append a few layers to the NN to handle control limits.}
    \label{fig:control_limit_concept}
\end{figure}

The key terms in \cref{thm:particular_xt} can be modified to account for control input constraints, as
\begin{align}
    \piUCL(\y_t) &= \textrm{Proj}_{\mathcal{U}_t}\left(\piclAu_{\ifacet,:,:}\y_t+ \piclbu_{:,\ifacet} \right) \label{eqn:piucl_control_constraints}\\
    \piLCL(\y_t) &= \textrm{Proj}_{\mathcal{U}_t}\left(\piclAl_{\ifacet,:,:}\y_t+ \piclbl_{:,\ifacet} \right) \label{eqn:pilcl_control_constraints},
\end{align}

A common example is box control input constraints.
The element-wise control input is,
\begin{align}
    \pi_{j,\ifacet}^{U_{CL}}(\y_t) =& \begin{cases}
        \mathrm{clip}(\pi_j^U(\y_t), \ubar{\mathbf{u}}_j, \bar{\mathbf{u}}_j), & \mathrm{if}\ \AoutjBtj \geq 0 \\
        \mathrm{clip}(\pi_j^L(\y_t), \ubar{\mathbf{u}}_j, \bar{\mathbf{u}}_j), & \mathrm{otherwise}
    \end{cases}
    \label{eqn:pi_upoly_elementwise_control_constraints},
\end{align}
where $\mathrm{clip}$ saturates the control if it exceeds the limits.
However, $\mathrm{clip}$ could be non-convex depending on the domain of $\x_t$ (and violates DCP rules in~\texttt{cvxpy}~\cite{diamond2016cvxpy} regardless), as visualized in~\cref{fig:control_limit_concept}.

We instead utilize the insights from~\cite{hu2020reach} that $\mathrm{clip}$ looks like two ReLUs stitched together (one is inverted).
In particular, we make the following changes to the NN:
\begin{enumerate}
    \item Subtract $\ubar{\mathbf{u}}_t$ from final layer's bias, $\mathbf{b}^{(m)} \leftarrow \mathbf{b}^{(m)} - \ubar{\mathbf{u}}_t$
    \item Append ReLU activation
    \item Append linear layer, $\mathbf{W}^{(m+1)}\shortleftarrow{0.2cm}-I_{n_u}, \mathbf{b}^{(m+1)}\shortleftarrow{0.2cm} \bar{\mathbf{u}}_t-\ubar{\mathbf{u}}_t$
    \item Append ReLU activation
    \item Append linear layer, $\mathbf{W}^{(m+2)}\leftarrow-I_{n_u}, \mathbf{b}^{(m+2)}\leftarrow\bar{\mathbf{u}}_t$.
\end{enumerate}
All of the preceding results apply to this modified NN for a system with box control input constraints.
Note that adding these ReLUs does not limit which types of activations can be used in other layers of the NN.

While these additional layers exactly mimic $\mathrm{clip}$ for any particular $\mathbf{y}_t$, they can induce additional conservatism because the additional ReLUs must be relaxed as well.
Additionally, while CROWN provides several options for the slope of the lower bound for a ReLU relaxation, note that the two additional ReLUs should be relaxed with a zero-slope lower bound to ensure no values exceed the control limits.

% \subsection{Uncertainty in Dynamics}\label{sec:forward_reachability:uncertainty_in_dynamics}

% Thus far, the system dynamics were assumed to contain known system matrices and known bounds on additive noise terms, as~\cref{eqn:closed_loop_dynamics}.
% In this section, we extend the approach to handle systems with uncertain matrices,
% \begin{align}
% \begin{split}
%     \mathbf{x}_{t+1} &= (A_t + \Tilde{A}_t) \mathbf{x}_{t} + B_t \mathbf{u}_t + \mathbf{c}_t + \bm{\omega}_t \label{eqn:ltv_dynamics_with_uncertainty} \\
%     \mathbf{y}_{t} &= \mathbf{C}_t^T\mathbf{x}_t + \bm{\nu}_t,
% \end{split}
% \end{align}
% where $A_t + \Tilde{A}_t \in [\ubar{A}_t, \, \bar{A}_t] \quad \forall t$.
% \mfe{Discuss uncertainty in $A$, how to handle (new lemma?)}
% \mfe{Add a plot with dynamics uncertainty.}

\subsection{Extension to Polynomial Dynamics}\label{sec:forward_reachability:polynomial}

This section extends the above work to systems with polynomial dynamics. Consider the system with dynamics \begin{align}
\begin{split}
    \mathbf{x}_{t+1} &= \mathbf{A}_t \mathbf{x}_{t} + \bar{f}(\mathbf{x}_t) + \mathbf{B}_t \mathbf{u}_t + \mathbf{c}_t + \bm{\omega}_t \label{eqn:ltv_dynamics_poly} \\
    \mathbf{y}_{t} &= \mathbf{C}_t^T\mathbf{x}_t + \bm{\nu}_t,
\end{split}
\end{align}
where $\bar{f}(\mathbf{x}):\R^n \to  \R^n$ is a polynomial and the other terms are defined in the same way as in \cref{eqn:ltv_dynamics}.
As a result, the counterpart of \cref{eqn:global_upper_bnd_generic_optimization} and \cref{eqn:global_lower_bnd_generic_optimization} in Lemma \ref{thm:bounds_on_xt1} are respectively as follows
\begin{align}
    \gamma_{t+1,\mathfrak{j}}^{U} &= \max_{\xinX} \Muj \mathbf{x}_t + f_\mathfrak{j}(\mathbf{x}_t) + \Nuj \label{eqn:global_upper_bnd_generic_optimization_poly} \\
    \gamma_{t+1,\mathfrak{j}}^{L} &= \min_{\xinX} \Mlj \mathbf{x}_t + f_\mathfrak{j}(\mathbf{x}_t) + \Nlj\label{eqn:global_lower_bnd_generic_optimization_poly},
\end{align}
where $f_\mathfrak{j}(\mathbf{x}_t)=\Aoutj \bar{f}(\mathbf{x}_t)$.
Because of the polynomial term $f_\mathfrak{j}(\mathbf{x}_t)$ in \cref{eqn:global_upper_bnd_generic_optimization_poly} and \cref{eqn:global_lower_bnd_generic_optimization_poly}, these objectives can be nonconvex and NP-Hard to solve for global optimality~\cite{parrilo2003minimizing}.
Thus, we adopt a principled convex relaxation technique~\cite{luo2010semidefinite} to compute a guaranteed upper and lower bound for \cref{eqn:global_upper_bnd_generic_optimization_poly} and \cref{eqn:global_lower_bnd_generic_optimization_poly}, respectively.

We first transform the polynomial term $f(\mathbf{x}_t)$ to a linear or quadratic term by introducing new variables $\mathbf{s}\in \R^{n_s}$ and new quadratic constraints.
For example, $f(\mathbf{x}_t) = \mathbf{x}_{t; 1}\mathbf{x}_{t; 2}\mathbf{x}_{t; 3}\mathbf{x}_{t; 4}$ is equivalent to $f(\mathbf{s})=\mathbf{s}_1\mathbf{s}_2$, with $\mathbf{s}_1=\mathbf{x}_{t; 1}\mathbf{x}_{t; 2}$ and $\mathbf{s}_2=\mathbf{x}_{t; 3}\mathbf{x}_{t; 4}$.
Then \cref{eqn:global_upper_bnd_generic_optimization_poly} can be transformed as a quadratically constrained quadratic programming (QCQP) in the following
\begin{align}
    \bm{\gamma}^{U} &= \max_{\mathbf{x}\in \mathcal{X}, \bar{\mathbf{s}}\in \mathcal{Q}} \Mu \mathbf{x} + \bar{\mathbf{s}}^T \mathbf{Q}_0\bar{\mathbf{s}} + \mathbf{b}_0^T\bar{\mathbf{s}} + \Nu \label{eqn:global_upper_bnd_generic_optimization_poly_qcqp} 
\end{align}
where the subscripts are dropped here for clarity, $\bar{\mathbf{s}}=[\mathbf{x}^T, \mathbf{s}^T]^T\in \R^{n+n_s}$, $\mathbf{Q}_0\in\mathcal{S}^{n+n_s}, \mathbf{b}_0\in \R^{n+n_s}$ and $\mathcal{Q}=\{\bar{\mathbf{s}}\ \lvert\ \bar{\mathbf{s}}^T \mathbf{Q}_i\bar{\mathbf{s}} + \mathbf{b}_i^T\bar{\mathbf{s}} + c_i = 0, \mathbf{Q}_i\in\mathcal{S}^{n+n_s}, \mathbf{b}_i\in \R^{n+n_s}, c_i\in \R, \forall i=1,\ldots\}$ is formulated by a sequence of quadratic constraints.
For QCQP, the semidefinite relaxation can give the tightest bounds among existing convex relaxation techniques~\cite{luo2010semidefinite}.
The semidefinite relaxation of \cref{eqn:global_upper_bnd_generic_optimization_poly_qcqp} can be formulated as  
\begin{align}
    \bm{\gamma}^{U} &= \max_{\mathbf{x}\in \mathcal{X}, \mathbf{S} \in \bar{\mathcal{Q}}, \mathbf{S}\succeq \textbf{0} } \Mu \mathbf{x} + \Tr(\bar{\mathbf{Q}}_0 \mathbf{S}) + \Nu \label{eqn:global_upper_bnd_generic_optimization_poly_qcqp_sdr} 
\end{align}
where $\mathbf{S}=[1,\bar{\mathbf{s}}^T]^T[1,\bar{\mathbf{s}}^T]\in \mathcal{S}^{1+n+n_s}$, $\bar{\mathbf{Q}}_i = [0, \mathbf{b}_i^T/2; \mathbf{b}_i/2, \mathbf{Q}_i], i=0,\ldots$ and $\bar{\mathcal{Q}} = \{\mathbf{S}\ \lvert\ \Tr(\bar{\mathbf{Q}}_i \mathbf{S}) + c_i =0,i=1,\ldots\}$. \cref{eqn:global_upper_bnd_generic_optimization_poly_qcqp_sdr} is equivalent to the original QCQP \cref{eqn:global_upper_bnd_generic_optimization_poly_qcqp} if an extra nonconvex constraint $\textbf{rank}(S) =1$ is added.
As a result, solving the convex program \cref{eqn:global_upper_bnd_generic_optimization_poly_qcqp_sdr} will yield a guaranteed upper bound for \cref{eqn:global_upper_bnd_generic_optimization_poly}.
The lower bound for \cref{eqn:global_lower_bnd_generic_optimization_poly} follows similarly.

\section{Backward Reachability Analysis}\label{sec:backward_reachability}

So far, this paper has considered the \textit{forward} reachability problem: starting from a set $\mathcal{X}_0$, what are all the possible states the system could occupy at time $t$?
In this section, we discuss the opposite problem of \textit{backward} reachability: given a target set $\mathcal{X}_T$, from which states will the system end up in the target set?
Depending on whether the initial or target state sets are known a priori and/or change frequently, one of these two paradigms would be a better fit for a given application.
Furthermore, the target set could represent a goal set or avoid/collision set -- either way, this section provides a framework for guaranteeing the system will reach/avoid the target set.

Traditionally, it is straightforward to switch between forward and backward reachability analysis through a change of variables~\cite{bansal2017hamilton,evans1998partial}.
However, including a NN in the analysis adds new challenges.
In particular, a key challenge with propagating sets \textit{backward} through NNs is that common NN activations have finite range (e.g., $\mathrm{ReLU}(x) = 0$ could correspond to any $x \leq 0$), which causes the sets to quickly explode to $\infty$.
Moreover, even using an infinite range activation (e.g., Leaky ReLU), the weight matrices may not be invertible (e.g., singular, insufficient rank).
Our approach is able to avoid these issues and still leverage forward propagation tools (e.g., CROWN~\cite{zhang2018efficient}) despite thinking backward in time.

Recent related work has developed specific invertible-by-design NN architectures~\cite{ardizzone2018analyzing} and modifications to the training procedure to regularize for invertibility~\cite{behrmann2019invertible}.
In contrast, our approach can be applied to any NN for which an affine relaxation can be computed, i.e., the same broad class of NN architectures (e.g., feedforward NNs with general activation functions) with arbitrary training processes that CROWN (or recent extensions~\cite{xu2020automatic,xu2021fast,wang2021beta}) operates on.

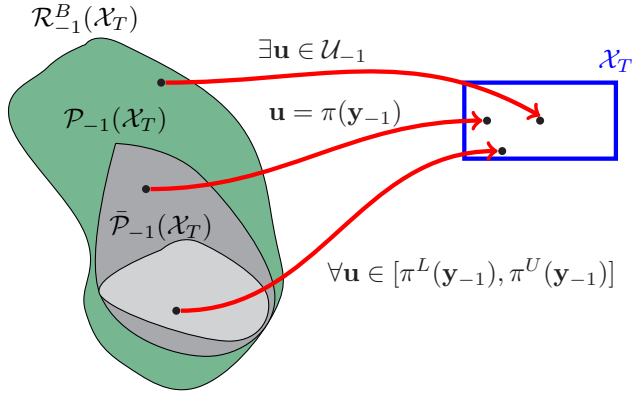
\begin{figure}[t]
\begin{tikzpicture}

    \draw [blue, ultra thick] (2,1) rectangle (4,2) node[above] {$\mathcal{X}_{T}$};
    
    \draw [fill=black!20!green!60!white]  plot[smooth, tension=.7] coordinates {(-3,2.5) (-3.5,2.2) (-4,1.5) (-3,0) (-3,-1) (-2.5,-1.7) (-1.5,-2) (-0.5,-1) (-0.5,0) (-1,1.5) (-2.5,2.5) (-3,2.5)} node[above] {$\mathcal{R}^B_{-1}(\mathcal{X}_T)$};
    \node[circle,fill=black,inner sep=0pt,minimum size=3pt] (a) at (-2,2) {};
    \node[circle,fill=black,inner sep=0pt,minimum size=3pt] (b) at (3,1.5) {};
    \draw [->,red, ultra thick] (a.east) to [out=0,in=150] (b.north);
    \node[black] (RTset) at (-0,2.4) {$\exists \mathbf{u} \in \mathcal{U}_{-1}$};

    \draw [fill=black!40!white]  plot[smooth, tension=.7] coordinates {(-2.6,1.2) (-1.5,0.8) (-0.5,-0.8) (-1.4,-1.6) (-2.8,-0.8) (-2.6,1.2)} node[above] {$\mathcal{P}_{-1}(\mathcal{X}_T)$};
    \node[circle,fill=black,inner sep=0pt,minimum size=3pt] (e) at (-2.2,0.6) {};
    \node[circle,fill=black,inner sep=0pt,minimum size=3pt] (f) at (2.3,1.5) {};
    \draw [->,red, ultra thick] (e.east) to [out=0,in=180] (f.west);
    \node[black] (PTset) at (0.3, 1.6) {$\mathbf{u} = \pi(\mathbf{y}_{-1})$};

    \draw [fill=black!20!white]  plot[smooth, tension=.7] coordinates {(-2,-0.2) (-1.5,-0.1) (-0.6,-0.8) (-1.1,-1.4) (-2.8,-0.8) (-2,-0.2)} node[above] {$\bar{\mathcal{P}}_{-1}(\mathcal{X}_T)$};
    \node[circle,fill=black,inner sep=0pt,minimum size=3pt] (c) at (-1.8,-1.0) {};
    \node[circle,fill=black,inner sep=0pt,minimum size=3pt] (d) at (2.5,1.1) {};
    \draw [->,red, ultra thick] (c.east) to [out=0,in=180] (d.west);
    \node[black] (PTbarset) at (2.1, -0.5) {$\forall \mathbf{u} \in [\pi^L(\mathbf{y}_{-1}), \pi^U(\mathbf{y}_{-1})]$};

\end{tikzpicture}
\caption{Backreachable, backprojection, and target sets. Given a target set, $\mathcal{X}_T$, the backreachable set $\mathcal{R}^B_{-1}(\mathcal{X}_T)$ contains all states for which \textit{some} control exists to move the system to $\mathcal{X}_T$ in one timestep. The backprojection set, $\mathcal{P}_{-1}(\mathcal{X}_T)$ contains all states for which the NN controller leads the system to $\mathcal{X}_T$. Inner-approximation $\bar{\mathcal{P}}_{-1}(\mathcal{X}_T)$ contains all states for which all controls that the relaxed NN could apply lead the system to $\mathcal{X}_T$.}
\label{fig:backreachable}
\end{figure}

\subsection{Backreachable \& Backprojection Sets}

\cref{fig:backreachable} illustrates the important concepts for this analysis.
Three different sets (left) contain states that lead to the target set $\mathcal{X}_T$ under different control inputs, as defined below.
The nomenclature of these sets is motivated by~\cite{lavalle2006planning}, but with slightly different definitions.

The first of these sets is the \textit{backward reachable set}, 
\begin{align}
% \begin{split}
    \mathcal{R}^B_{t-1}(\mathcal{X}_T) &= \{ \mathbf{x}\ \lvert\ \exists\mathbf{u} \in \mathcal{U}_{t-1} \mathrm{\ s.t.\ } \label{eqn:backreachable_sets} \\
     &\mathbf{A}_{t-1}\mathbf{x} + \mathbf{B}_{t-1}\mathbf{u} + \mathbf{c}_{t-1} \in \mathcal{R}^B_t(\mathcal{X}_T) \}\ \forall t \leq 0  \nonumber \\
    \mathcal{R}^B_0(\mathcal{X}_T)&=\mathcal{X}_T, \nonumber
% \end{split}
\end{align}
which consists of states $\mathbf{x}_t$ for which some admissible control trajectory, $\mathbf{u}_{t:T}$, takes the system to $\mathcal{X}_T$.
To simplify the notation, we assume no noise (i.e., $\ubar{\bm{\omega}}_t = \bar{\bm{\omega}}_t = \ubar{\bm{\nu}}_t = \bar{\bm{\nu}}_t = \mathbf{0}$) and perfect observations, $\mathbf{C}^T_t = \mathbf{I} \Rightarrow \mathbf{y}_t = \mathbf{x}_t$ in this section.
Backward reachable sets are particularly useful when the controller is truly a ``black box'' or has not yet been defined (since the definition only uses the fact that $\mathbf{u}\in\mathcal{U}$, rather than $\pi$).

In the case of a trained NN control policy, we can be more precise about the closed loop system's safety properties.
In particular, for the closed-loop system $f$ from~\cref{eqn:closed_loop_dynamics}, we denote $\mathcal{P}_{-t}(\mathcal{X}_T)$ the \textit{backprojection set}, which contains all states that will lead to a given target set $\mathcal{X}_T\subseteq\R^{n_x}$ in exactly $t$ timesteps, as defined by the recursion
\begin{align}
% \begin{split}
    \mathcal{P}_{t-1}(\mathcal{X}_T) &= \{ \mathbf{x}\ \lvert\ \mathbf{A}_{t-1}\mathbf{x} + \mathbf{B}_{t-1}\pi(\mathbf{x}) + \mathbf{c}_{t-1} \in \mathcal{P}_t(\mathcal{X}_T) \} \label{eqn:backprojection_sets} \nonumber \\
    & \quad\quad \forall t \leq 0 \nonumber \\
    \mathcal{P}_0(\mathcal{X}_T)&=\mathcal{X}_T.
% \end{split}
\end{align}

Recall that in~\cref{sec:background:reach_avoid_verification_problem}, we computed over-approximations of the reachable sets because the NN makes exact reachability analysis computationally challenging.
Analogously, we will compute \textit{under}-approximations of the backprojection sets, $\bar{\mathcal{P}}_{-t}(\mathcal{X}_T)\subseteq\mathcal{P}_{-t}(\mathcal{X}_T)$.
Under-approximations are still useful because if $\bar{\mathcal{P}}_{-t}(\mathcal{X}_T)\supseteq \mathcal{X}_0$, then $\mathcal{P}_{-t}(\mathcal{X}_T)\supseteq \mathcal{X}_0$ as well, meaning for every state in $\mathcal{X}_0$, the control policy $\pi$ is guaranteed to lead the system to $\mathcal{X}_T$ in $T$ timesteps.
Thus, it is desirable for $\bar{\mathcal{P}}_{-t}(\mathcal{X}_T)$ to be as close to $\mathcal{P}_{-t}(\mathcal{X}_T)$ as possible.

\subsection{Deriving Backprojection Sets}

We first provide a summary of the proposed approach:
\begin{enumerate}
    \item Ignoring the NN, find (hyper-)rectangular bounds on the backreachable set (this is a superset of the backprojection set associated with the particular NN controller)
    \item Within this region, relax the NN controller to acquire upper/lower affine bounds on control
    \item Compute the states that will lead to the target set for every control effort within the upper/lower bounds
\end{enumerate}
This last step provides an under-approximation of the backprojection set, which is the set of interest.

The following lemma provides polytope bounds on $\bar{\mathcal{P}}_{t}(\mathcal{X}_T)$ for a single timestep, which is the key component of the recursive algorithm introduced in the next section.

\begin{lemma}\label[lemma]{thm:backprojection}
Given an $m$-layer NN control policy $\pi:\R^{n_y}\to\R^{n_u}$, closed-loop dynamics $f: \R^{n_x} \times \Pi \to \R^{n_x}$ as in~\cref{eqn:ltv_dynamics,eqn:closed_loop_dynamics}, and bounds on the state set, $\ubar{\mathbf{x}}_{t+1} \leq \mathbf{x}_{t+1} \leq \bar{\mathbf{x}}_{t+1}$, the following polytope bounds on the previous state set hold:
\begin{align}
\backprojA \mathbf{x}_t \leq \backprojb, \label{eqn:backreachable:polytope}
\end{align}
where $\mathbf{Z}^{\mathbf{B}\CROWNAl\CROWNAu}, \mathbf{Z}^{\mathbf{B}\CROWNAu\CROWNAl} \in \R^{n_x \times n_y}$ and $\mathbf{z}^{\mathbf{B}\CROWNbl\CROWNbu}, \mathbf{z}^{\mathbf{B}\CROWNbu\CROWNbl} \in \R^{n_x}$ are defined $\forall \istate \in [n_x]$ (by row/element),
\begin{align}
\left[\mathbf{Z}^{\mathbf{B}\CROWNAl\CROWNAu}\right]_{\istate,:} &= \mathbf{B}_{t;\istate,:} \Jbar{\mathbf{B}_{t; \istate, :}}{\CROWNAl}{\CROWNAu} \label{eqn:backreachable:zba}\\
\left[\mathbf{Z}^{\mathbf{B}\CROWNAu\CROWNAl}\right]_{\istate,:} &= \mathbf{B}_{t;\istate,:} \Jbar{\mathbf{B}_{t; \istate, :}}{\CROWNAu}{\CROWNAl} \label{eqn:backreachable:zbb} \\
\left[\mathbf{z}^{\mathbf{B}\CROWNbl\CROWNbu}\right]_{\istate} &= \mathbf{B}_{t;\istate,:} \Jbar{\mathbf{B}_{t; \istate, :}}{\CROWNbl}{\CROWNbu} \label{eqn:backreachable:zbc} \\
\left[\mathbf{z}^{\mathbf{B}\CROWNbu\CROWNbl}\right]_{\istate} &= \mathbf{B}_{t;\istate,:} \Jbar{\mathbf{B}_{t; \istate, :}}{\CROWNbu}{\CROWNbl}, \label{eqn:backreachable:zbd}
\end{align}
and $\ubar{\mathbf{x}}_t, \bar{\mathbf{x}}_t$ are computed elementwise by solving the LPs in~\cref{eqn:backreachable:lpmax_obj,eqn:backreachable:lpmax_constr_state,eqn:backreachable:lpmax_constr_control,eqn:backreachable:lpmin_obj,eqn:backreachable:lpmin_constr_state,eqn:backreachable:lpmin_constr_control}.

\begin{proof}
Given dynamics from~\cref{eqn:ltv_dynamics,eqn:closed_loop_dynamics}, solve the following optimization problems for each state $\istate \in [n_x]$,
\begin{align}
\bar{\mathbf{x}}_{t; \istate} = \max_{\mathbf{x}_t, \mathbf{u}_t} \quad & \mathbf{x}_{t; \istate} \label{eqn:backreachable:lpmax_obj} \\
\mathrm{s.t.} \quad & \mathbf{A}_t \mathbf{x}_t + \mathbf{B}_t \mathbf{u}_t + \mathbf{c}_t \in \mathcal{X}_{t+1} \label{eqn:backreachable:lpmax_constr_state} \\
& \mathbf{u}_t \in \mathcal{U}_t \label{eqn:backreachable:lpmax_constr_control} \\
\ubar{\mathbf{x}}_{t; \istate} = \min_{\mathbf{x}_t, \mathbf{u}_t} \quad & \mathbf{x}_{t; \istate} \label{eqn:backreachable:lpmin_obj} \\
\mathrm{s.t.} \quad & \mathbf{A}_t \mathbf{x}_t + \mathbf{B}_t \mathbf{u}_t + \mathbf{c}_t \in \mathcal{X}_{t+1} \label{eqn:backreachable:lpmin_constr_state} \\
& \mathbf{u}_t \in \mathcal{U}_t, \label{eqn:backreachable:lpmin_constr_control}
\end{align}
which provides a (hyper-)rectangular outer bound ($\ubar{\mathbf{x}}_{t} \leq \mathbf{x}_{t} \leq \bar{\mathbf{x}}_{t}$) on the \textit{backreachable set}.
Note that this is a LP for convex $\mathcal{X}_{t+1}, \mathcal{U}_t$ as we will use here.

Given that bound, $\ubar{\mathbf{x}}_{t} \leq \mathbf{x}_{t} \leq \bar{\mathbf{x}}_{t}$, \cref{thm:crown_particular_x} provides $\CROWNAu, \CROWNAl, \CROWNbu, \CROWNbl$ which lead to the following inequalities based on~\cref{thm:particular_xt} with $\Aout=\mathbf{I}_{n_x}$, $\forall \istate \in [n_x]$,
\begin{align}
\begin{split}
& \left[ \mathbf{A}_t \mathbf{x}_t + \mathbf{B}_t ( \Jbar{\mathbf{B}_{t; \istate, :}}{\CROWNAl}{\CROWNAu} \mathbf{y}_t + \Jbar{\mathbf{B}_{t; \istate, :}}{\CROWNbl}{\CROWNbu}) + \mathbf{c}_t \right]_{\istate} \\
 \leq\ 
& \mathbf{x}_{t+1; \istate} \\
 \leq\ 
& \left[ \mathbf{A}_t \mathbf{x}_t + \mathbf{B}_t ( \Jbar{\mathbf{B}_{t; \istate, :}}{\CROWNAu}{\CROWNAl} \mathbf{y}_t + \Jbar{\mathbf{B}_{t; \istate, :}}{\CROWNbu}{\CROWNbl}) + \mathbf{c}_t \right]_{\istate}.
\end{split}
\end{align}

After substituting in for $\mathbf{y}_t$, grouping $\mathbf{x}_t$ terms together, and using \cref{eqn:backreachable:zba,eqn:backreachable:zbb,eqn:backreachable:zbc,eqn:backreachable:zbd},
\begin{align}
\begin{split}
& \left(\mathbf{A}_{t} + \mathbf{Z}^{\mathbf{B}\CROWNAl\CROWNAu}\right) \mathbf{x}_{t} + \mathbf{z}^{\mathbf{B}\CROWNbl\CROWNbu} + \mathbf{c}_t \\
 \leq\ 
& \mathbf{x}_{t+1} \\
 \leq\ 
& \left(\mathbf{A}_{t} + \mathbf{Z}^{\mathbf{B}\CROWNAu\CROWNAl}\right) \mathbf{x}_{t} + \mathbf{z}^{\mathbf{B}\CROWNbu\CROWNbl} + \mathbf{c}_t
\label{eqn:backproj:relaxed_nn_bounds}.
\end{split}
\end{align}

To ensure no NN control pushes the system into a state beyond $[\ubar{\mathbf{x}}_{t+1}, \bar{\mathbf{x}}_{t+1}]$, the following inequalities hold,
\begin{align}
\ubar{\mathbf{x}}_{t+1} \leq \left(\mathbf{A}_{t} + \mathbf{Z}^{\mathbf{B}\CROWNAl\CROWNAu}\right) \mathbf{x}_{t} + \mathbf{z}^{\mathbf{B}\CROWNbl\CROWNbu} + \mathbf{c}_t \label{eqn:backproj:lower_next_state} \\
\left(\mathbf{A}_{t} + \mathbf{Z}^{\mathbf{B}\CROWNAu\CROWNAl}\right) \mathbf{x}_{t} + \mathbf{z}^{\mathbf{B}\CROWNbu\CROWNbl} + \mathbf{c}_t \leq \bar{\mathbf{x}}_{t+1} \label{eqn:backproj:upper_next_state} .
\end{align}

Written in matrix form,
\begin{align}
\begin{bmatrix}
\mathbf{A}_{t} + \mathbf{Z}^{\mathbf{B}\CROWNAu\CROWNAl} \\
- \left(\mathbf{A}_{t} + \mathbf{Z}^{\mathbf{B}\CROWNAl\CROWNAu}\right)
\end{bmatrix} \mathbf{x}_{t} \leq
\begin{bmatrix}
\bar{\mathbf{x}}_{t+1} - \left( \mathbf{z}^{\mathbf{B}\CROWNbu\CROWNbl} + \mathbf{c}_t \right) \\
- \ubar{\mathbf{x}}_{t+1} + \mathbf{z}^{\mathbf{B}\CROWNbl\CROWNbu} + \mathbf{c}_t
\end{bmatrix}. \label{eqn:backreachable:polytope_only_dyn}
\end{align}

To arrive at~\cref{eqn:backreachable:polytope}, append $\ubar{\mathbf{x}}_{t} \leq \mathbf{x}_{t} \leq \bar{\mathbf{x}}_{t}$ to \cref{eqn:backreachable:polytope_only_dyn}.
\end{proof}
\end{lemma}

\begin{corollary}
The set $\bar{\mathcal{P}}_{-1}(\mathcal{X}_T)$ composed of all $\mathbf{x}_t$ that satisfy~\cref{eqn:backreachable:polytope} is a subset of the one-step backprojection set, $\mathcal{P}_{-1}(\mathcal{X}_T)$.

\begin{proof}
Suppose $\mathbf{x}_t$ satisfies \cref{eqn:backreachable:polytope}, i.e., $\mathbf{x}_t \in \bar{\mathcal{P}}_{-1}(\mathcal{X}_T)$.
Thus, $\mathbf{x}_t$ also satisfies the inequalities~\cref{eqn:backproj:lower_next_state,eqn:backproj:upper_next_state}, and thus also \cref{eqn:backproj:relaxed_nn_bounds}.
Since $\mathbf{x}_{t+1}=f(\mathbf{x}_t; \pi)$, $\mathbf{x}_t$ also satisfies $\ubar{\mathbf{x}}_{t+1} \leq f(\mathbf{x}_t; \pi) \leq \bar{\mathbf{x}}_{t+1}$, which means that $\mathbf{x}_t \in \mathcal{P}_{-1}(\mathcal{X}_T)$ as well.
Thus, $\bar{\mathcal{P}}_{-1}(\mathcal{X}_T) \subseteq \mathcal{P}_{-1}(\mathcal{X}_T)$.
\end{proof}
\end{corollary}

\subsection{Algorithm for Computing Backprojection Sets}

\begin{algorithm}[t]
 \caption{Backprojection Set Estimation}
 \begin{algorithmic}[1]
 \setcounter{ALC@unique}{0}
 \renewcommand{\algorithmicrequire}{\textbf{Input:}}
 \renewcommand{\algorithmicensure}{\textbf{Output:}}
 \REQUIRE target state set $\mathcal{X}_T$, trained NN control policy $\pi$, time horizon $\tau$, partition parameter $\mathbf{r}$
 \ENSURE backprojection set approximations $\bar{\mathcal{P}}_{-\tau:0}(\mathcal{X}_T)$
    \STATE $\bar{\mathcal{P}}_{0}(\mathcal{X}_T) \leftarrow \mathcal{X}_T$ \label{alg:backprojection:initialize}
    \FOR{$t$ in $\{-1, -2, \ldots, -\tau\}$} \label{alg:backprojection:timestep_for_loop}
        \STATE $\bar{\mathcal{P}}_{t}(\mathcal{X}_T) \leftarrow \emptyset$
        \STATE $\bar{\mathcal{R}}^B_{t}(\mathcal{X}_T) = [\ubar{\ubar{\mathbf{x}}}_{t}, \bar{\bar{\mathbf{x}}}_{t}] \leftarrow \mathrm{backreach}(\bar{\mathcal{P}}_{t+1}(\mathcal{X}_T), \mathcal{U}_{t})$ \label{alg:backprojection:backreach} \\
        \STATE $\mathcal{S} \leftarrow \mathrm{partition}([\ubar{\ubar{\mathbf{x}}}_{t}, \bar{\bar{\mathbf{x}}}_{t}], \mathbf{r})$
        \FOR{$[\ubar{\mathbf{x}}_{t}, \bar{\mathbf{x}}_{t}]$ in $\mathcal{S}$}
            \STATE $\CROWNAu, \CROWNAl, \CROWNbu, \CROWNbl \leftarrow \mathrm{CROWN}(\pi, [\ubar{\mathbf{x}}_{t}, \bar{\mathbf{x}}_{t}])$
            % \STATE $\mathbf{Z}, \mathbf{z} \leftarrow \mathrm{range\_to\_polytope}([\ubar{\mathbf{x}}_{t}, \bar{\mathbf{x}}_{t}]_i)$
            \STATE $\mathbf{P} \leftarrow \backprojA$ \label{alg:backprojection:polytope_constraint_A}
            \vspace{0.02in}
            \STATE $\mathbf{p} \leftarrow \backprojb$ \label{alg:backprojection:polytope_constraint_b}
            \vspace{0.02in}
            \STATE $\mathcal{A} \leftarrow \{\mathbf{x}\ \lvert\ \mathbf{P} \mathbf{x} \leq \mathbf{p} \}$
            \STATE $\bar{\mathcal{P}}_{t}(\mathcal{X}_T) \leftarrow \bar{\mathcal{P}}_{t}(\mathcal{X}_T) \cup \mathcal{A}$ \label{alg:backprojection:union}
        \ENDFOR
    \ENDFOR
 \RETURN $\bar{\mathcal{P}}_{-\tau:0}(\mathcal{X}_T)$ \label{alg:backprojection:return}
 \end{algorithmic}\label{alg:backprojection}
\end{algorithm}

\cref{thm:backprojection} provides a formula for computing $\bar{\mathcal{P}}_{-1}(\mathcal{X}_T)$, i.e., an under-approximation of the backprojection set for a single timestep.
This section extends that idea to enable computing $\bar{\mathcal{P}}_{-\tau:0}(\mathcal{X}_T)$ for a time horizon $\tau$ and also describes how partitioning can help improve the results, particularly when \cref{eqn:backreachable:polytope} is the empty set.
The proposed procedure is summarized in~\cref{alg:backprojection}.

After initializing the zeroth backprojection set as the target set (\cref{alg:backprojection:initialize}), we recursively compute $\bar{\mathcal{P}}_{t}(\mathcal{X}_T)$ backward in time (\cref{alg:backprojection:timestep_for_loop}).
For each timestep, we first compute an over-approximation on the backreachable set, $\bar{\mathcal{R}}^B_{t}(\mathcal{X}_T)$, by solving the LPs in \cref{eqn:backreachable:lpmax_obj,eqn:backreachable:lpmax_constr_control,eqn:backreachable:lpmax_constr_state} and \cref{eqn:backreachable:lpmin_obj,eqn:backreachable:lpmin_constr_control,eqn:backreachable:lpmin_constr_state} (\cref{alg:backprojection:backreach}).

Since $\mathcal{R}^B_{t}(\mathcal{X}_T) \supseteq \mathcal{P}_{t}(\mathcal{X}_T)$, we can relax the NN over $\mathcal{R}^B_{t}(\mathcal{X}_T)$, and that relaxation will also hold for all states in $\mathcal{P}_{t}(\mathcal{X}_T)$ and $\bar{\mathcal{P}}_{t}(\mathcal{X}_T)$.
However, $\mathcal{R}^B_{t}(\mathcal{X}_T)$ could be large, which could lead to a loose NN relaxation.
Thus, instead of analyzing $\mathcal{R}^B_{t}(\mathcal{X}_T)$ as a single set, we uniformly partition the set $\mathcal{R}^B_{t}(\mathcal{X}_T)$ into $\mathbf{r}\in\mathds{N}^{n_x}$ cells (defined per dimension).

For each cell in the partition, we have a (hyper-)rectangular set of states, $[\ubar{\mathbf{x}}_{t}, \bar{\mathbf{x}}_{t}]_i$, that could be passed as inputs to the NN controller.
The NN can be relaxed over this set using CROWN (\cref{thm:crown_particular_x}) or a similar type of algorithm.
In particular, we extract $\CROWNAu, \CROWNAl, \CROWNbu, \CROWNbl$ from CROWN, which define $\mathbf{P}$ and $\mathbf{p}$, the constraints in \cref{eqn:backreachable:polytope} (\cref{alg:backprojection:polytope_constraint_A,alg:backprojection:polytope_constraint_b}).
These constraints define a polytope of states that are guaranteed to lead to $\bar{\mathcal{P}}_{t+1}(\mathcal{X}_T)$ at the next timestep (for any control within CROWN's bounds), so those states should be added to $\bar{\mathcal{P}}_{t}(\mathcal{X}_T)$ (\cref{alg:backprojection:union}).
Note that it is possible, despite partitioning, that $\mathcal{A}=\emptyset$.
After following this procedure for each timestep, the algorithm returns a sequence of inner-approximations of the backprojection sets (\cref{alg:backprojection:return}).

%!TEX root=main.tex

\section{Experimental Results}

\begin{table}[t]
\centering
\small{
		\begin{tabular}{|c|c|c|}
			\hline
			 Algorithm                    & Runtime [s]         &   Error \\
			\hline
             CL-CROWN                            & $0.023 \pm 0.000$ &  654556 \\
             CL-SG-IBP~\cite{xiang2020reachable} & $0.888 \pm 0.011$ &   70218 \\
            \hline
			 Reach-SDP~\cite{hu2020reach} & $42.57 \pm 0.54$  &     207 \\
			 Reach-SDP-Partition          & $670.26 \pm 2.91$ &      12 \\
			 Reach-LP                     & $0.017 \pm 0.000$   &    1590 \\
			 Reach-LP-Partition           & $0.263 \pm 0.001$   &      34 \\
			\hline
			% \hline
			% Orig Results... & & \\
			% Algorithm & Runtime [s] \mfe{Re-run!} & Error \\
			% \hline
			% Reach-SDP~\cite{hu2020reach} & 20.31 & 206 \\ 
			% Reach-SDP-Partition & 347.14 & \textbf{19.35} \\ 
			% Reach-LP & \textbf{0.63}& 848 \\ 
			% Reach-LP-Partition & 9.87 & 19.87 \\ 
			% \hline
	    \end{tabular}
	    }
\caption{Reachable Sets for Double Integrator. The first two methods analyze the NN and dynamics separately, leading to high error accumulation, while the next 4 methods analyze the NN and dynamics together. Reach-LP is $4,000\times$ faster to compute but $7\times$ looser than Reach-SDP~\cite{hu2020reach}. Reach-LP-Partition refines the Reach-LP bounds by splitting the input set into 16 subsets, giving $150\times$ faster computation time and $5\times$ tighter bounds than Reach-SDP~\cite{hu2020reach}.}
\label{tab:double_integrator_reachable_set}
\end{table}

\begin{figure}[t]
	\centering
	\begin{subfigure}{0.5\linewidth}
		\centering
		\includegraphics[width=\linewidth]{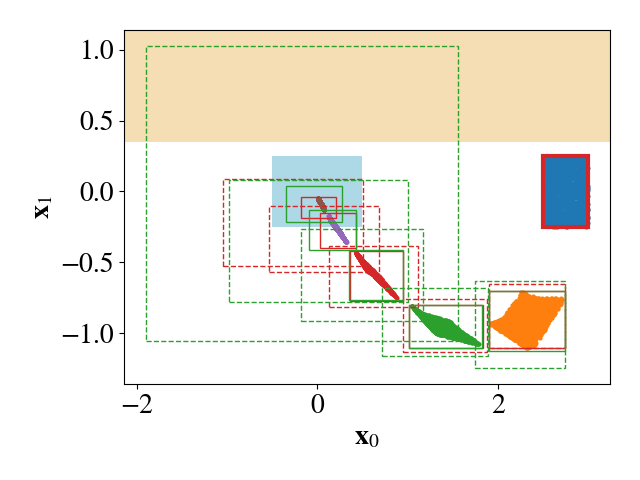}
		\captionsetup{justification=centering}
		\caption{Reachable Set Estimates}
		\label{fig:double_integrator_reachable_set_trajectory}
	\end{subfigure}%
	\begin{subfigure}{0.5\linewidth}
	    \includegraphics[width=\linewidth]{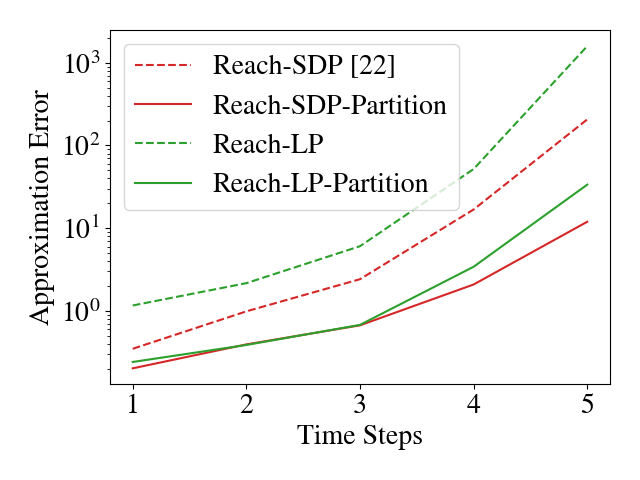}
	    \captionsetup{justification=centering}
	    \caption{Over-approximation error}
		\label{fig:double_integrator_reachable_set_error}
	\end{subfigure}%
	\caption{Reachable Sets for Double Integrator. In (a), all reachable set algorithms bound sampled states across the timesteps, starting from the blue $\mathcal{X}_0$, and the tightness of these bounds is quantified per timestep in (b).}
	\label{fig:double_integrator_reachable_set}
	\vspace{-0.2in}
\end{figure}

This section demonstrates our convex reachability analysis tool, Reach-LP, on simulated scenarios.
We first show an example verification task and quantify the improvement in runtime vs. bound tightness over the state-of-the-art~\cite{hu2020reach} for a double integrator system.
We then apply the algorithm on a 6D quadrotor model subject to multiple sources of noise.

\subsection{Double Integrator}\label{sec:results:double_integrator}

Consider the LTI double integrator system from~\cite{hu2020reach}, 
\begin{equation}
    \mathbf{x}_{t+1}=\underbrace{\begin{bmatrix}
        1 & 1 \\ 0 & 1
    \end{bmatrix}}_{\mathbf{A}_t} \mathbf{x}_t + \underbrace{\begin{bmatrix}
        0.5 \\ 1
    \end{bmatrix}}_{\mathbf{B}_t} \mathbf{u}_t,
\end{equation}
with $\mathbf{c}_t=0$, $\mathbf{C}_t=\mathbf{I}_2$ and no noise, discretized with sampling time $t_s=1s$.
As in~\cite{hu2020reach}, we implemented a linear MPC with prediction horizon $N_{MPC}=10$, weighting matrices $\mathbf{Q}=\mathbf{I}_2, \mathbf{R}=1$, and terminal weighting matrix $\mathbf{P}_{\infty}$ synthesized from the discrete-time Riccati equation, subject to state constraints $\mathcal{A}^C=[-5,5]\times[-1,1]$ and input constraint $\mathbf{u}_t\in[-1,1]\forall t$.
We used MPC to generate 2420 samples of state and input pairs then trained a NN with Keras~\cite{chollet2015keras} for 20 epochs with batch size 32.

\subsection{Comparison with Baseline}\label{sec:results:comparison_with_reachsdp}

\cref{tab:double_integrator_reachable_set} and \cref{fig:double_integrator_reachable_set} compare several algorithms on the double integrator system using a NN with [5,5] neurons and ReLU activations\footnote{The proposed methods apply similarly to NNs with other activation functions, including sigmoid and tanh}.
The key takeaway is that Reach-LP-Partition provides a $5\times$ improvement in reachable set tightness over the prior state-of-the-art, Reach-SDP~\cite{hu2020reach} (which does not use input set partitioning), while requiring $150\times$ less computation time.
We implemented Reach-SDP in Python with \texttt{cvxpy} and \texttt{MOSEK}~\cite{andersen2000mosek}.
All computation times (until \cref{sec:results:high_dim}) are reported from an i7-6700K CPU with 16GB RAM on Ubuntu 20.04.

We also note the large error accumulation that occurs when the NN and dynamics are analyzed separately, as in \cite{xiang2020reachable,yang2019efficient}, as that assumes the NN could output its extreme values from any state in $\mathcal{X}_t$ (top 2 rows of \cref{tab:double_integrator_reachable_set}).
CL-CROWN is a simple baseline where, at each timestep, the NN output bounds $\mathcal{U}^\text{NN}_t$ are computed with CROWN~\cite{zhang2018efficient}, then \cref{eqn:nfl_analysis_exact} is solved (in closed-form) assuming $u_t\in\mathcal{U}^\text{NN}_t$.
Similarly, CL-SG-IBP is the method from~\cite{xiang2020reachable}, where $\mathcal{U}^\text{NN}_t$ is instead computed with SimGuided-IBP to resolution $\varepsilon=0.1$.

\cref{fig:double_integrator_reachable_set_trajectory} shows sampled trajectories, where each colored cluster of points represents sampled reachable states at a particular timestep (blue$\rightarrow$orange$\rightarrow$green, etc.).
Recall that sampling trajectories could miss possible reachable states, whereas these algorithms are guaranteed to over-approximate the reachable sets.
Reachable set bounds are visualized for various algorithms: Reach-SDP~\cite{hu2020reach}, Reach-LP, and those two algorithms after partitioning the input set into 16 cells.
The key takeaway is that while all approaches provide outer bounds on the sampled trajectories, the algorithms provide various degrees of tightness to the sampled points.

We quantify tightness as the ratio of areas between the smallest axis-aligned bounding box on the sampled points and the provided reachable set (minus 1), shown in \cref{fig:double_integrator_reachable_set_error} as the system progresses forward in time.
Note that as expected, all algorithms get worse as the number of timesteps increase, but that Reach-LP-Partition and Reach-SDP-Partition perform the best and similarly.
This provides numerical comparisons of the rectangle sizes from~\cref{fig:double_integrator_reachable_set_trajectory}.

Note that both Reach-LP and Reach-SDP methods could be improved by properly choosing the direction of polytope facets.
Additionally, while Reach-SDP can provide ellipsoidal bounds given the quadratic nature of the formulation, we implement only the polytope bounds in this comparison.

\subsection{Verification}\label{sec:results:verification}

A primary application of reachable sets is to verify reach-avoid properties.
In~\cref{fig:double_integrator_reachable_set_trajectory}, we consider a case with an avoid set $\mathcal{A}=\{\x\ \lvert\  \x_1\geq0.35\}$ (orange) and a goal set $\mathcal{G}=[-0.5,0.5]\times[-0.25,0.25]$ (cyan).
Reach-LP and Reach-SDP-Partition, verify these properties for this 5-step scenario, highlighting the importance of tight reachable sets.

\subsection{Scalability to Deep NNs}\label{sec:results:scalability}

\begin{figure}
\centering
	\begin{subfigure}{0.54\linewidth}
	\centering
		\includegraphics[width=\linewidth, trim =20 20 40 20, clip]{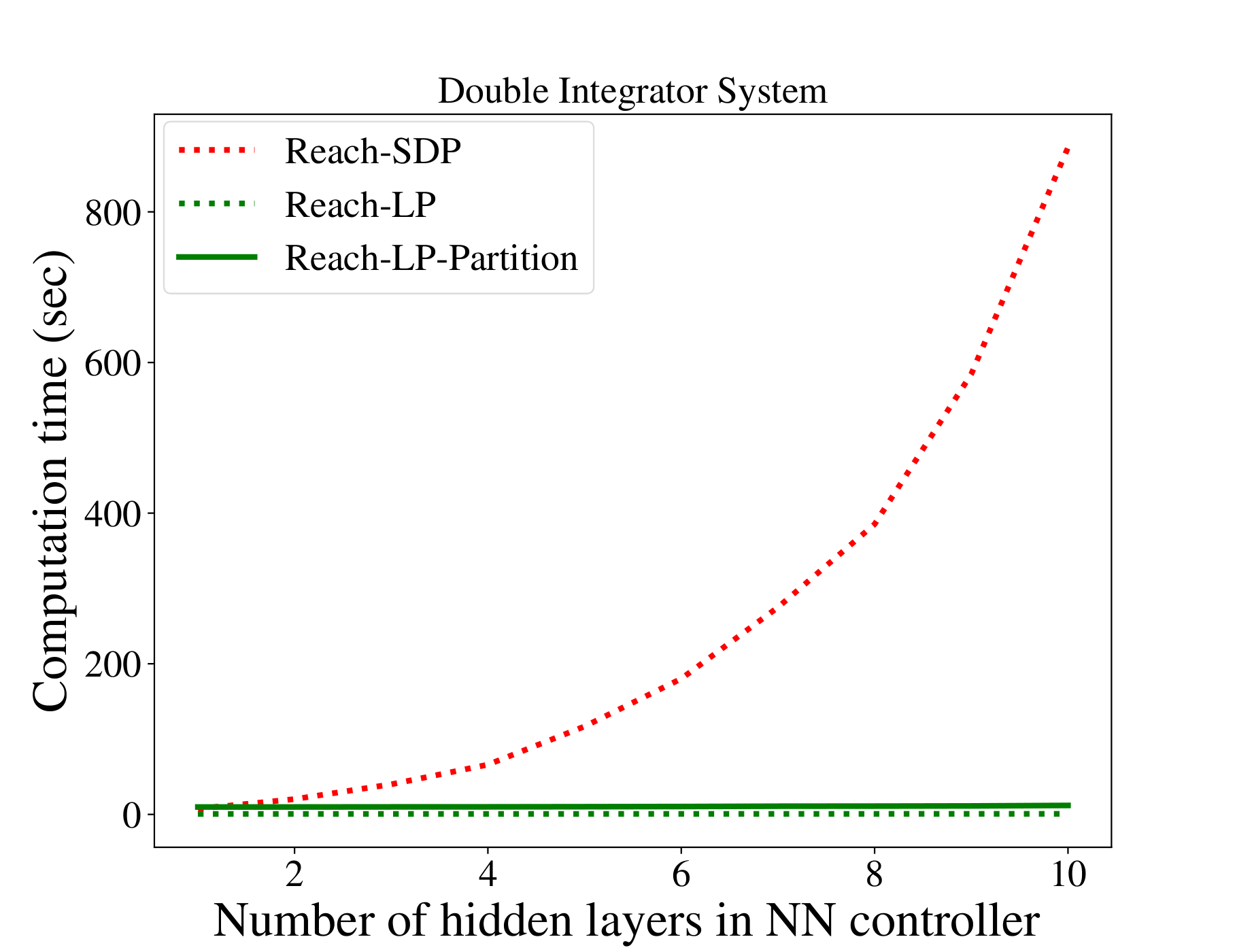}
		\caption{Runtime with NN depth}
	\label{fig:time_vs_num_layers}
	\end{subfigure}%
	\begin{subfigure}{0.46\linewidth}
	\centering
\includegraphics[width=\linewidth, trim =20 15 20 30, clip]{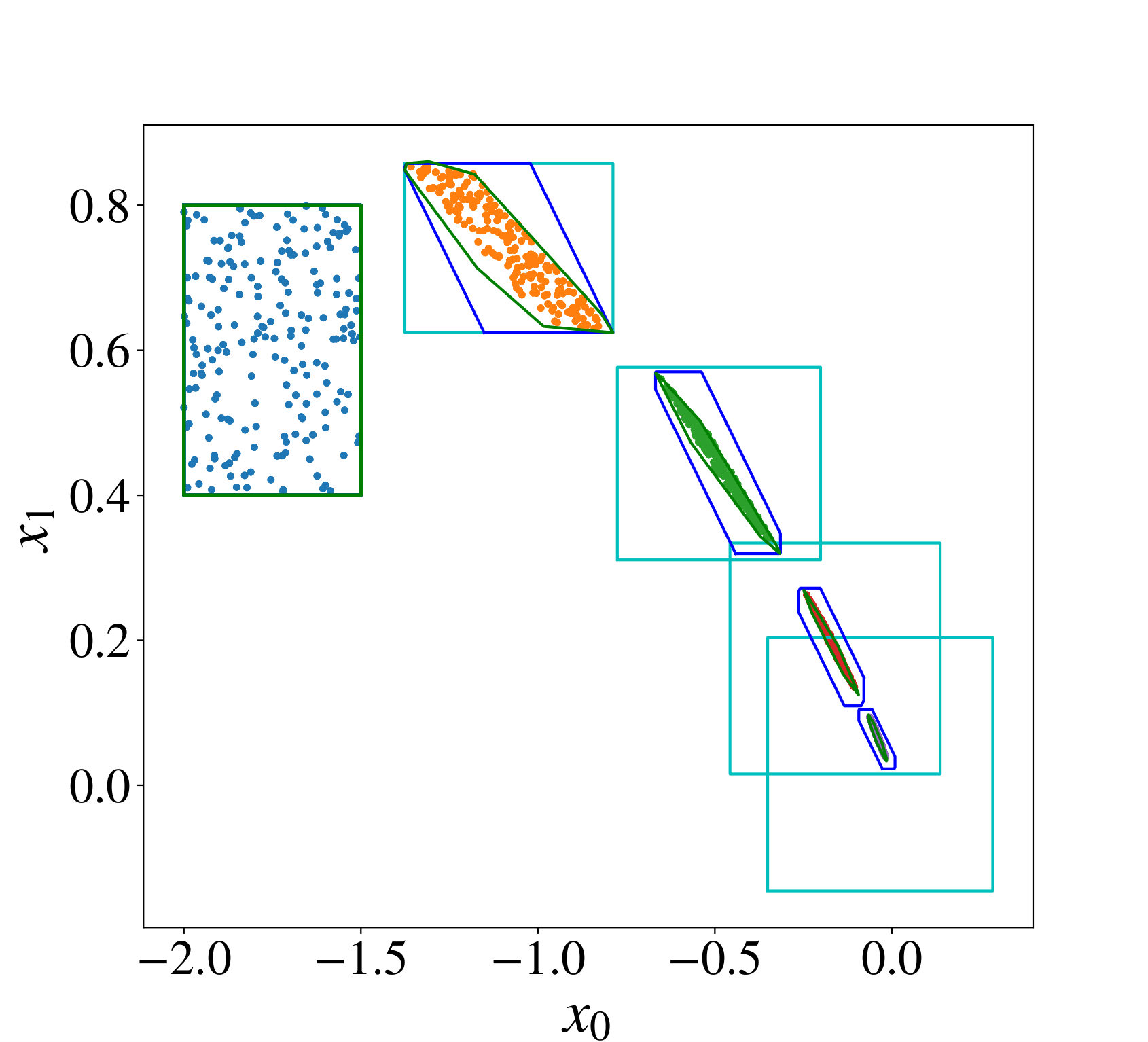}
    \caption{Number of Polytope Facets}
	\label{fig:double_integrator_reachable_set_polytope_facets}	
	\end{subfigure}
	\caption{(a) Our linear relaxation-based methods (Reach-LP, Reach-LP-Partition) scale well for deeper NNs (Reach-LP: 0.6 to 0.74s), whereas SDP-based methods grow to intractable runtimes. Note that input set partitioning multiplies computation time by a scalar.(b)  Using Reach-LP, the bounding shapes correspond to $l_{\infty}$-ball, 8-Polytope, and 35-Polytope. Reachable sets become tighter with more facets.}
	% \caption{(a) Our linear relaxation-based methods (Reach-LP, Reach-LP-Partition) scale well for deeper NNs (Reach-LP: 0.6 to 0.74s), whereas SDP-based methods grow to intractable runtimes. Note that input set partitioning multiplies computation time by a scalar.(b)  Using Reach-LP, the bounding shapes correspond to \textcolor{Cyan}{$l_{\infty}$-ball}, \textcolor{blue}{8-Polytope}, and \textcolor{ForestGreen}{35-Polytope}. Reachable sets become tighter with more facets.}
	\label{fig:double_integrator_result}
\end{figure}

To demonstrate the scalability of the method, we trained NNs with 1-10 hidden layers of 5 neurons and report the average runtime of 5 trials of reachability analysis of the double integrator system.
In~\cref{fig:time_vs_num_layers}, while Reach-SDP appears to grow exponentially (taking $>800s$ for a 10-layer NN), our proposed Reach-LP methods remain very efficient ($<0.75s$ for Reach-LP on all NNs).
Note that we omit Reach-SDP-Partition ($\sim16\times$ more than Reach-SDP) from this plot to maintain reasonable scale.
Recall that CROWN itself has $O(m^2 n^3)$ time complexity~\cite{zhang2018efficient}, for an $m$-layer network with $n$ neurons per layer and $n$ outputs.

\subsection[Ablation Study: Lp-balls vs. Polytopes]{Ablation Study: $\ell_{\infty}$ vs. Polytopes}\label{sec:results:ablation_lp_vs_polytope}

Recall that~\cref{sec:forward_reachability} described reachable sets as either polytopes or $\ell_\infty$-balls.
\cref{fig:double_integrator_reachable_set_polytope_facets} shows the effect of that choice: as the number of sides of the polytope increases, the reachable set size decreases.
The tradeoff is that the computation time scales linearly with the number of sides on the polytope.
Note that a $\ell_\infty$-ball is a 4-polytope, and that $\mathcal{X}_0$ was chosen to show a different scenario than~\cref{fig:double_integrator_reachable_set}.

\subsection{Ablation Study: Closed-Form vs. LP}

\begin{table}[t]
\begin{subfigure}{\linewidth}
\centering
    \begin{tabular}{|c||c|c|c|}
\hline
 & \multicolumn{3}{c|}{\# Partitions} \\
 Solver & 1 & 4 & 16               \\
\hline
 L.P. & $0.229 \pm 0.018$ & $0.856 \pm 0.046$ & $3.308 \pm 0.091$ \\
 C.F. & $0.017 \pm 0.000$ & $0.066 \pm 0.000$ & $0.265 \pm 0.002$ \\
\hline
\end{tabular}
\caption{Double Integrator (5 timesteps)}
\end{subfigure}
\begin{subfigure}{\linewidth}
\vspace{0.1in}
\centering
\begin{tabular}{|c||c|c|c|}
\hline
 & \multicolumn{3}{c|}{\# Partitions} \\
 Solver & 1 & 4 & 16               \\
\hline
 L.P. & $2.375 \pm 0.039$ & $9.424 \pm 0.121$ & $37.932 \pm 0.508$ \\
 C.F. & $0.110 \pm 0.002$ & $0.432 \pm 0.001$ & $1.738 \pm 0.003$  \\
\hline
\end{tabular}
\caption{6D Quadrotor (2 seconds)}
\end{subfigure}
\caption{Runtime in seconds shows over an order of magnitude speed-up by using closed-form (C.F.) instead of linear program (L.P.) to solve~\cref{eqn:global_upper_bnd_generic_optimization} when applicable (i.e., $\ell_p$-ball sets).}
\label{tab:runtime_closed_form}
\end{table}

The computational benefit of the closed-form solution in~\cref{thm:closed_form} is shown in~\cref{tab:runtime_closed_form}.
For both the double integrator (a) and 6D quadrotor (b), the runtime of the closed-form (C.F.) solution is over an order of magnitude faster than with the LP solver.
This speedup is observed across various partition resolutions, with times reported as the mean $\pm$ one standard deviation of 5 repeats.
Note that the C.F. and LP solvers return the same reachable sets, so there is no tightness tradeoff -- solely a computational speedup.

\subsection{Greedy Sim-Guided Partitioning}

\begin{figure*}[t]
    \centering

    \begin{subfigure}{0.33\linewidth}
        \centering
        \includegraphics[width=\textwidth, trim=40 20 20 60, clip]{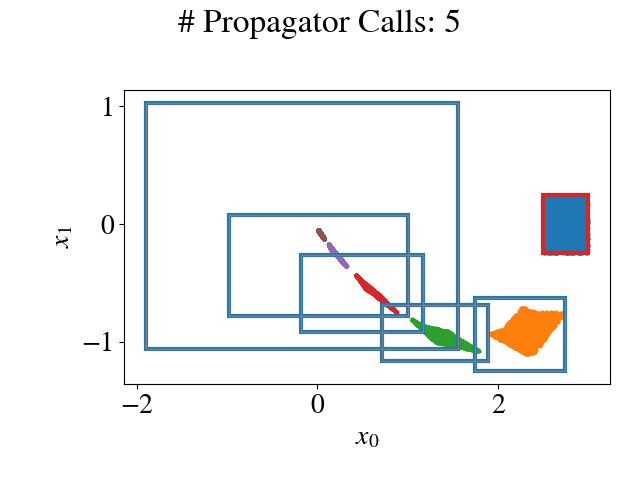}
        \captionsetup{justification=centering}
        \caption{5 Propagator Calls}
        \label{fig:partition_sequence:0}
    \end{subfigure}%
    \hfill
    \begin{subfigure}{0.33\linewidth}
        \centering
        \includegraphics[width=\textwidth, trim=40 20 20 60, clip]{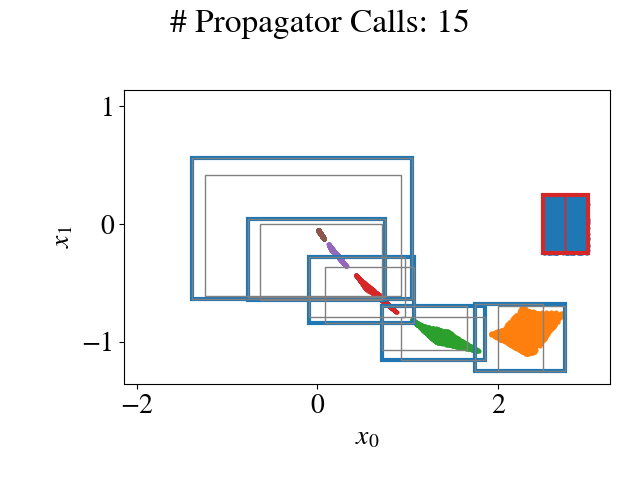}
        \captionsetup{justification=centering}
        \caption{15 Propagator Calls}
        \label{fig:partition_sequence:1}
    \end{subfigure}%
    \hfill
    \begin{subfigure}{0.33\linewidth}
        \centering
        \includegraphics[width=\textwidth, trim=40 20 20 60, clip]{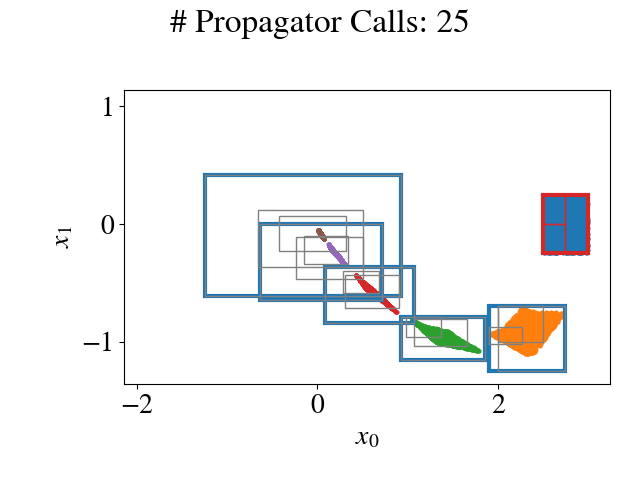}
        \captionsetup{justification=centering}
        \caption{25 Propagator Calls}
        \label{fig:partition_sequence:2}
    \end{subfigure}

    \vspace{0.1in}

    \begin{subfigure}{0.33\linewidth}
        \centering
        \includegraphics[width=\textwidth, trim=40 20 20 60, clip]{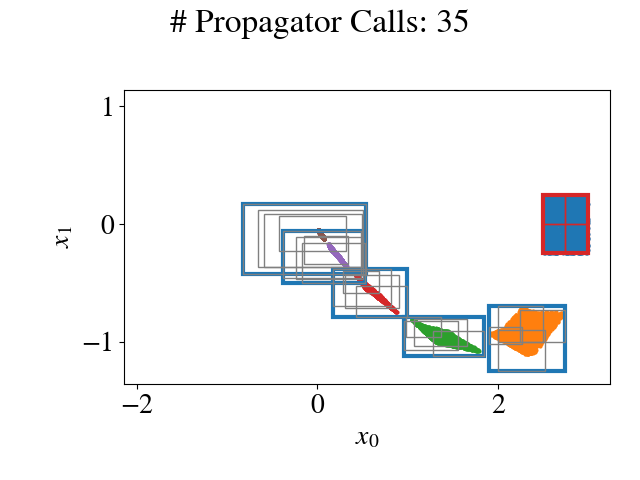}
        \captionsetup{justification=centering}
        \caption{35 Propagator Calls}
        \label{fig:partition_sequence:3}
    \end{subfigure}%
    \hfill
    \begin{subfigure}{0.33\linewidth}
        \centering
        \includegraphics[width=\textwidth, trim=40 20 20 60, clip]{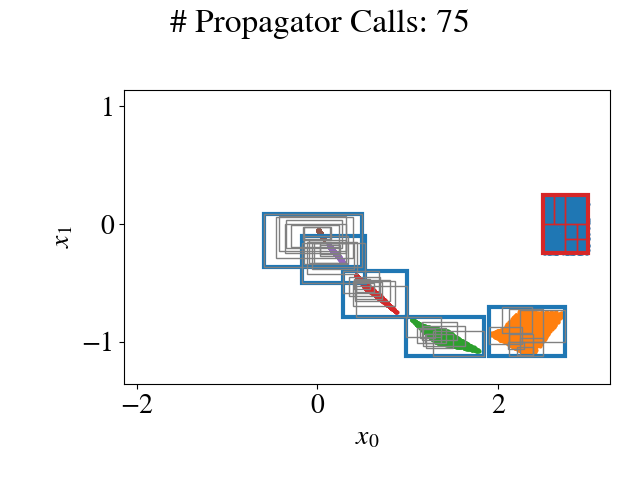}
        \captionsetup{justification=centering}
        \caption{75 Propagator Calls}
        \label{fig:partition_sequence:4}
    \end{subfigure}%
    \hfill
    \begin{subfigure}{0.33\linewidth}
        \centering
        \includegraphics[width=\textwidth, trim=40 20 20 60, clip]{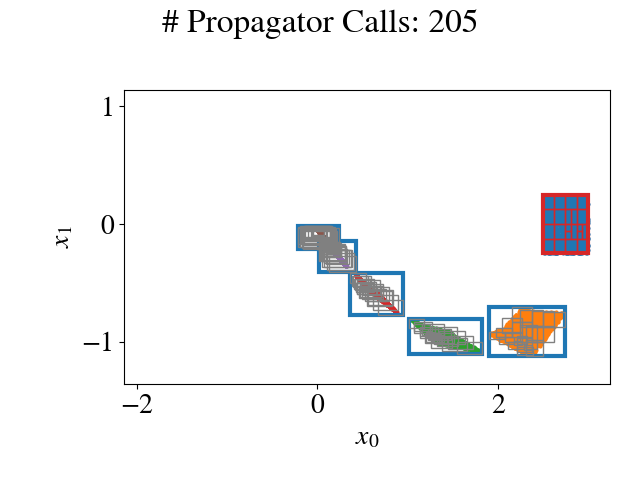}
        \captionsetup{justification=centering}
        \caption{205 Propagator Calls}
        \label{fig:partition_sequence:5}
    \end{subfigure}%

    \caption{Reach-LP with Closed-Loop Greedy Simulation-Guided Partitioning. As $\mathcal{X}_0$ (red box) is partitioned, the reachable set estimates (blue boxes) tighten to the Monte Carlo samples. Gray boxes show the reachable sets corresponding to cells of the $\mathcal{X}_0$ partition.}
    \label{fig:partition_sequence}
\end{figure*}

\cref{fig:partition_sequence} shows the refinement of reachable sets under the Closed-Loop Greedy Sim-Guided partitioner and CL-CROWN propagator.
The blue reachable set estimates become much tighter to the Monte Carlo samples, while still providing a guaranteed over-approximation of the true system's forward reachable sets.
Note that the partition of $\mathcal{X}_0$ is not uniform, as seen most clearly in \cref{fig:partition_sequence:4}.

\subsection{6D Quadrotor with Noise}\label{sec:results:quadrotor}

\begin{figure}
	\centering
	\begin{subfigure}{0.5\linewidth}
		\centering
		\includegraphics[width=\textwidth, trim =70 10 50 40, clip]{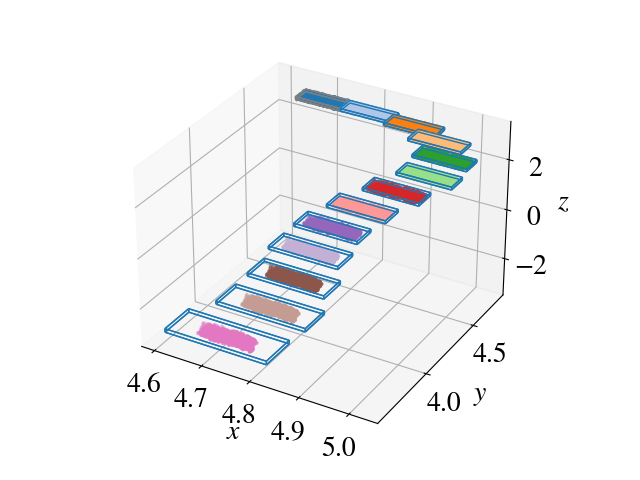}
		\captionsetup{justification=centering}
		\caption{No Noise}
		\label{fig:quadrotor_without_noise}
	\end{subfigure}%
	\begin{subfigure}{0.5\linewidth}
		\centering
		\includegraphics[width=\textwidth, trim =70 10 50 40, clip]{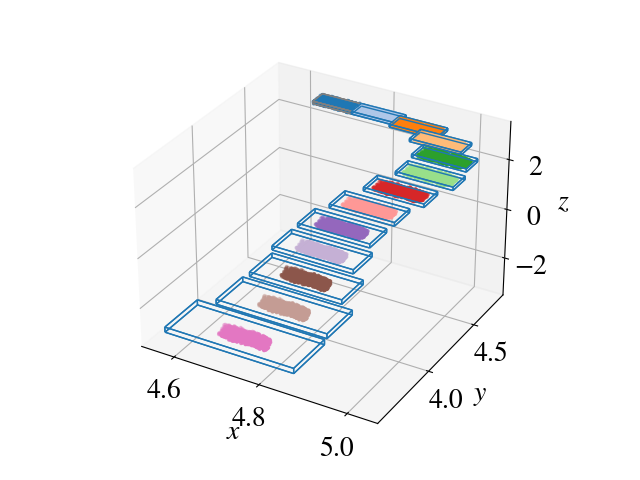}
		\captionsetup{justification=centering}
		\caption{Sensor \& Process Noise}
		\label{fig:quadrotor_with_noise}
	\end{subfigure}%
	\caption{Reachable Sets for 6D Quadrotor. Only $(x,y,z)$ states are shown, even though the reachable sets are computed in 6D. Blue boxes (Reach-LP) bound the clusters of sampled points at each discrete timestep, starting from the blue $\mathcal{X}_0$. It took $0.11$ sec to compute the 12 reachable sets per scenario. In (b), $\bm{\nu}\sim\texttt{Unif}(\pm0.001\cdot\mathds{1}_6), \bm{\omega}\sim\texttt{Unif}(\pm0.005\cdot\mathds{1}_6)$.}
	\label{fig:quadrotor}
	\vspace{-0.2in}
\end{figure}

Consider the 6D nonlinear quadrotor from~\cite{hu2020reach,lopez2019verification}, 
\begin{align}
    \dot{\mathbf{x}}&=\underbrace{\begin{bmatrix}
        0_{3\times3} & I_3 \\ 0_{3\times3} & 0_{3\times3}
    \end{bmatrix}}_{\mathbf{A}_t} \mathbf{x}_t + \underbrace{\begin{bmatrix}
         & g & 0 & 0 \\
        0_{3\times3} & 0 & -g & 0 \\
         & 0 & 0 & 1
    \end{bmatrix}^T}_{\mathbf{B}_t} \underbrace{\begin{bmatrix}
        \text{tan}(\theta) \\
        \text{tan}(\phi) \\
        \tau
    \end{bmatrix}}_{\mathbf{u}_t} \nonumber\\ &\quad\quad+
    \underbrace{\begin{bmatrix}
        0_{5\times1} \\
        -g
    \end{bmatrix}}_{\mathbf{\mathbf{c}}_t} + \bm{\omega}_t, \label{eqn:quadrotor_dynamics}
\end{align}
which differs from~\cite{hu2020reach,lopez2019verification} in that we add $\bm{\omega}_t$ as a uniform process noise, and that the output is measured as in~\cref{eqn:ltv_dynamics} with $\mathbf{C}_t=\mathbf{I}_6$, subject to uniform sensor noise.
As in~\cite{hu2020reach}, the state vector contains 3D positions and velocities, $[p_x,p_y,p_z,v_x,v_y,v_z]$, while nonlinearities from~\cite{lopez2019verification} are absorbed into the control as functions of $\theta$ (pitch), $\phi$ (roll), and $\tau$ (thrust) (subject to the same actuator constraints as~\cite{hu2020reach}).
We implemented a similar nonlinear MPC as~\cite{hu2020reach} in MATLAB to collect $(\mathbf{x}_t,\mathbf{u}_t)$ training pairs, then trained a [32,32] NN with Keras as above.
We use Euler integration to account for \cref{eqn:quadrotor_dynamics} in our discrete time formulation.

\cref{fig:quadrotor} shows the reachable sets with and without noise.
Note that while these plots only show $(x,y,z)$ position, the reachable sets are estimated in all 6D.
The first key takeaway is that the blue boxes (Reach-LP with $\ell_\infty$-balls) provide meaningful bounds for a long horizon (12 steps, $1.2s$ shown).
Secondly, unlike Reach-SDP, Reach-LP is guaranteed to bound worst-case noise realizations.
Computing these bounds took a total of $0.11 \pm 0.003$ seconds.

\subsection{High-Dimensional System}\label{sec:results:high_dim}

\begin{figure}[t]
\centering
\includegraphics[scale=0.5]{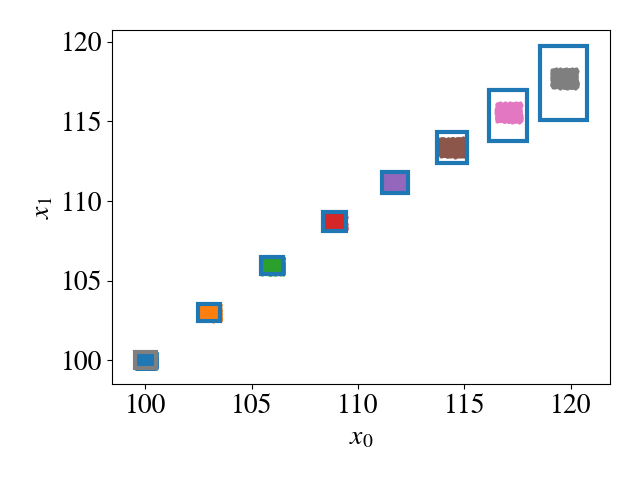}
\caption{Reachable Sets for the International Space Station (ISS) dynamics model with 270 states and a NN controller. Reach-LP scales to this high-dimensional system.}
\label{fig:iss}
\end{figure}

To demonstrate that Reach-LP scales to high-dimensional systems, \cref{fig:iss} uses the International Space Station (ISS) Component 1R model\footnote{http://slicot.org/20-site/126-benchmark-examples-for-model-reduction}, which has $n_x=270$ and $n_u = 3$.
After training a NN in the same manner as~\cref{sec:results:double_integrator}, reachable sets for all 270 states were computed, with only $(\mathbf{x}_0, \mathbf{x}_1)$ shown. The average time for each reachability analysis is about 25.34 sec on a desktop computer with i7-9700K CPU@3.60GHz (8 cores) and 32 GB RAM.

\subsection{Polynomial Dynamics}

\begin{figure*}[t]
	\centering
	\begin{subfigure}{0.25\linewidth}
		\centering
		\includegraphics[width=\textwidth, trim =20 20 20 20, clip]{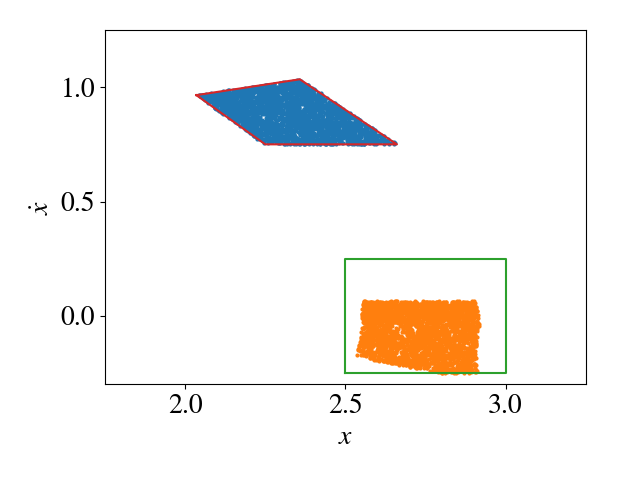}
		\captionsetup{justification=centering}
		\caption{$2 \times 2$ partition}
		\label{fig:backproj_22}
	\end{subfigure}%
	\begin{subfigure}{0.25\linewidth}
		\centering
		\includegraphics[width=\textwidth, trim =20 20 20 20, clip]{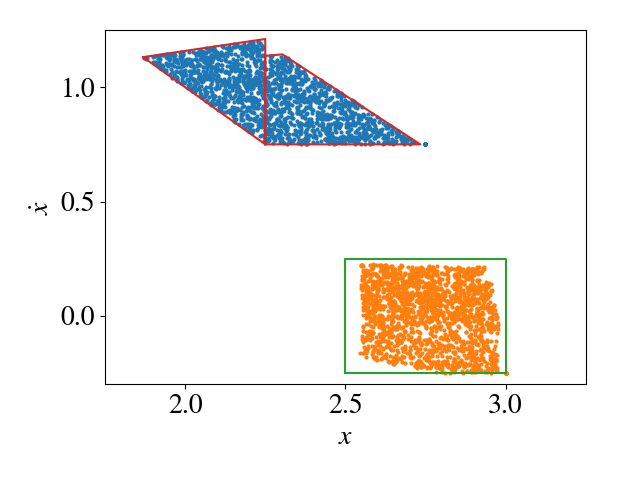}
		\captionsetup{justification=centering}
		\caption{$4 \times 4$ partition}
		\label{fig:backproj_44}
	\end{subfigure}%
	\begin{subfigure}{0.25\linewidth}
		\centering
		\includegraphics[width=\textwidth, trim =20 20 20 20, clip]{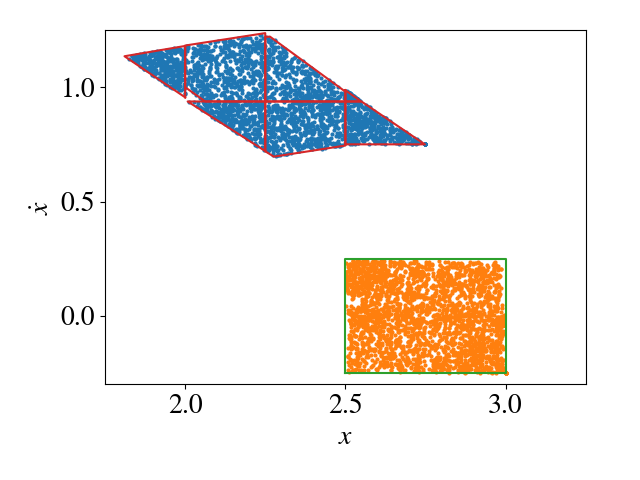}
		\captionsetup{justification=centering}
		\caption{$8 \times 8$ partition}
		\label{fig:backproj_88}
	\end{subfigure}%
	\begin{subfigure}{0.25\linewidth}
		\centering
		\includegraphics[width=\textwidth, trim =20 20 20 20, clip]{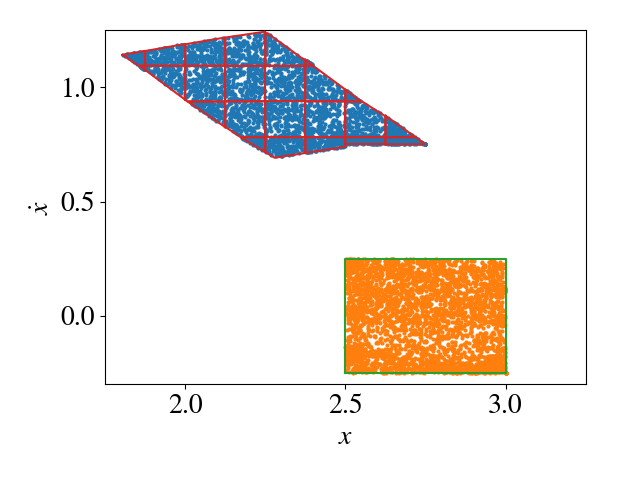}
		\captionsetup{justification=centering}
		\caption{$16 \times 16$ partition}
		\label{fig:backproj_1616}
	\end{subfigure}%
	\caption{1-step Backprojection Sets for Double Integrator. The green rectangle represents $\mathcal{X}_T$, which is partitioned into an $N \times N$ grid in (a)-(d) for various $N$.
	The estimated backprojection set of each cell of $\mathcal{X}_T$ is computed and visualized as a single red polytope (when non-empty).
	Thus, the estimated backprojection set of $\mathcal{X}_T$ is the union of all red polytopes.
	To confirm the results, blue points are sampled from the estimated backprojection sets, and the system state at the next timestep (under the NN control policy) are shown in orange. The orange points do indeed lie within $\mathcal{X}_T$, with better coverage of $\mathcal{X}_T$ as the number of partitions increases.}
	\label{fig:backproj}
\end{figure*}

\begin{figure}[t]
\centering
\includegraphics[scale=0.5]{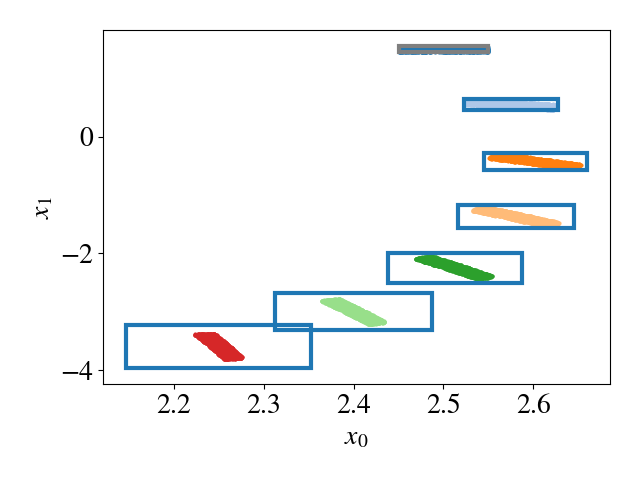}
\caption{Reachable Sets for the Duffing oscillator. Starting from $\mathcal{X}_0$ (gray rectangle, blue samples), the Reach-LP extension from~\cref{sec:forward_reachability:polynomial} is used to estimate reachable sets for 6 timesteps into the future, despite the nonlinear (polynomial) terms in the dynamics.}
\label{fig:duffing}
\end{figure}

To show the SDP-based method from~\cref{sec:forward_reachability:polynomial} on a system with polynomial dynamics, we consider the Duffing oscillator with the following dynamics
\begin{flalign}
    \dot{\mathbf{x}}_1 &= \mathbf{x}_2, \\
    \dot{\mathbf{x}}_2 &= - \mathbf{x}_1 - 2\zeta \mathbf{x}_2  - \mathbf{x}_1^3 + u,
\end{flalign}
where $\mathbf{x}\in\mathbb{R}^2$ and $u\in\mathbb{R}$ are the state and control inputs, respectively, and $\zeta=0.3$ is the damping coefficient.
Furthermore, $\mathbf{c}_t=0$, $\mathbf{C}_t=\mathbf{I}_2$, zero noise is considered, and the dynamics are discretized with sampling time $t_s=0.3s$.
We fit a neural network controller described in~\cite{dang2012reachability}.
Applying the techniques introduced in \cref{sec:forward_reachability:polynomial}, the reachable sets are illustrated in \cref{fig:duffing}.
The estimated reachable sets contain the sampled states, but due to the relaxation, the bounds of the reachable sets are relatively loose compared to the linear systems that did not require relaxations for the dynamics.

\subsection{Backward Reachability}

For the double integrator system and NN described in~\cref{sec:results:double_integrator}, \cref{fig:backproj} shows one-step backprojection sets for a given target set, $\mathcal{X}_T$.
Recall that this NN is not invertible, both because it uses ReLU activations and some weight matrix dimensions prevent inversion.

The green rectangle represents $\mathcal{X}_T = [2.5, 3.0] \times [-0.25, 0.25]$.
$\mathcal{X}_T$ is partitioned into a uniform grid of various resolution in (a)-(d), ranging from $2\times2$ to $16\times16$.
The estimated backprojection set of each cell is computed and visualized as a red polytope.
Thus, the estimated backprojection set of $\mathcal{X}_T$ is the union of all red polytopes.
In practice, if $\mathcal{X}_T$ represents a goal set, one could check that the system's starting state is inside one of the gray polytopes before ``pressing go'' on the system.

To demonstrate that points inside the backprojection set will move the system to the target set under the NN control policy, we uniformly sampled points in each polytope of the estimated backprojection set and simulated the system forward one step (using the NN control policy).
In~\cref{fig:backproj}, the blue points represent these samples at timestep $T-1$, and the orange points are the resulting system state at timestep $T$.
All of the orange points lie within the green rectangle of $\mathcal{X}_T$, which suggests the backprojection sets were computed properly.

Note that in (a), the under-approximation of the backprojection set has some clear gaps.
There are a few reasons for this, but the key issue is that $\mathcal{A}=\emptyset$ (from \cref{alg:backprojection}) for some of the cells of the $\mathcal{X}_T$ partition.
What causes $\mathcal{A}=\emptyset$?
If $\mathcal{X}_T$ is large, the backward reachable set could be as well.
Thus, even a partitioned $\mathcal{X}_T$ could lead to a large backward reachable set partition, which is used as the NN input set for CROWN.
If the input set to CROWN is large, the NN relaxations can be relatively loose.
Loose bounds on the NN restrict the size of $\bar{\mathcal{P}}$, since $\bar{\mathcal{P}}$ only includes states for which that \textit{entire} range of relaxed control inputs will lead to $\mathcal{X}_T$ (see bottom arrow of \cref{fig:backreachable}).
To get around this issue, (b)-(d) repeat the analysis with finer partitions, which substantially improves the coverage of samples across $\mathcal{X}_T$.
Future work could investigate better methods for partitioning $\mathcal{X}_T$ (e.g., using simulation as guidance~\cite{xiang2020reachable,everett2020robustness}).

%!TEX root=main.tex

\section{Future Directions}

Many open research directions remain in analyzing NFLs.
We first describe several natural extensions of this work.
One existing challenge is in mitigating the conservatism due to the accumulation of approximation error over many timesteps.
This could be addressed, for example, by replacing the recursive calculations with a method for computing $n$-step reachable sets at once.
In addition, capturing other types of uncertainties and nonlinearities (e.g., uncertainty in $\mathbf{A}_t$ and $\mathbf{B}_t$ matrices) will expand the set of neural feedback loops that can be analyzed.
Continuous time systems, which likely have hybrid dynamics due to the NN control policy, present additional opportunities for expanding the framework beyond the Euler integration approach described here.

More broadly, an open challenge is in synthesizing provably robust control policies.
While this work focused on analyzing NFLs with a pre-trained NN, future work could explore modifications to the training process to enable faster or tighter online analysis.
Moreover, bringing similar ideas to stability analysis could provide additional notions of robustness beyond reachability.

Finally, the adoption of these analysis methods on real safety-critical systems will require realistic measurement/perception models, as many modern systems use high-dimensional sensors (e.g., camera, lidar), which are often fed directly into perception NNs (e.g., for object detection or segmentation).
The effects of uncertainty propagating through these modular pipelines presents new challenges before such systems are ready to be deployed in safety-critical settings, such as robots operating alongside humans.

\section{Conclusion}\label{sec:conclusion}

This paper proposed a convex relaxation-based algorithm for computing forward reachable sets and backprojection sets of NFLs, which are closed-loop systems with NN controllers.
Prior work is limited to shallow NNs and is computationally intensive, which limits applicability to real systems.
Furthermore, our method accounts for measurement of sensor and process noise, as well as nonlinearities in the dynamics.
The results show that this work advances the state-of-the-art in guaranteeing properties of systems that employ NNs in the feedback loop.

% \clearpage
\bibliographystyle{ieeetr} 
\bibliography{main}

\begin{IEEEbiography}[{\includegraphics[width=1in,height=1.25in,clip,keepaspectratio]{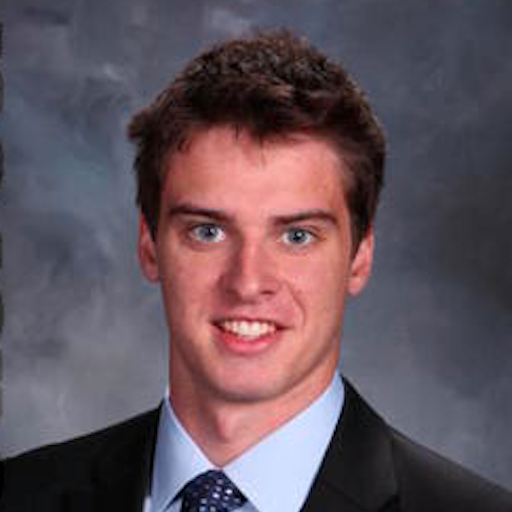}}]{Michael Everett}
received the S.B., S.M., and Ph.D. degrees in mechanical engineering from the Massachusetts Institute of Technology (MIT), in 2015, 2017, and 2020, respectively. He is currently a Research Scientist with the Department of Aeronautics and Astronautics at MIT. His research lies at the intersection of machine learning, robotics, and control theory, with specific interests in the theory and application of safe and robust neural feedback loops. He was an author of works that won the Best Paper Award on Cognitive Robotics at IROS 2019, the Best Student Paper Award and a Finalist for the Best Paper Award on Cognitive Robotics at IROS 2017, and a Finalist for the Best Multi-Robot Systems Paper Award at ICRA 2017. He has been interviewed live on the air by BBC Radio and his team’s robots were featured by Today Show and the Boston Globe.
\end{IEEEbiography}

\begin{IEEEbiography}[{\includegraphics[trim=50 0 50 0,width=1in,clip,keepaspectratio]{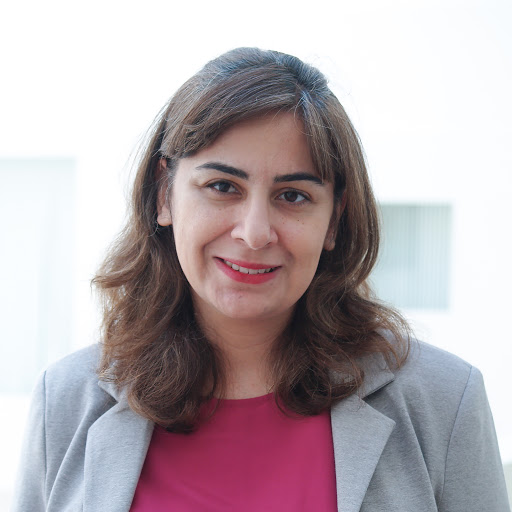}}]{Golnaz Habibi}  is currently a Research Scientist with the Department of Aeronautics and Astronautics at MIT. Golnaz received her Bsc in electrical and control engineering from K.N.Toosi University of Technology, Iran, in 2005, her Msc. in control engineering from Tarbiat Modares University, Iran, in 2007, and her PhD in computer science from Rice University, in 2015. Golnaz is broadly interested in robotics, control systems, machine learning, and multi agent systems. Her current research focuses on visual navigation, reliable communication, and improving the safety and reliability of autonomous agents. Her paper has been nominated for best student paper award in DARS 2012 and she received the K2I award by Chevron Co. in 2013.
\end{IEEEbiography}
\begin{IEEEbiography}[{\includegraphics[width=1in,height=1.25in,clip,keepaspectratio]{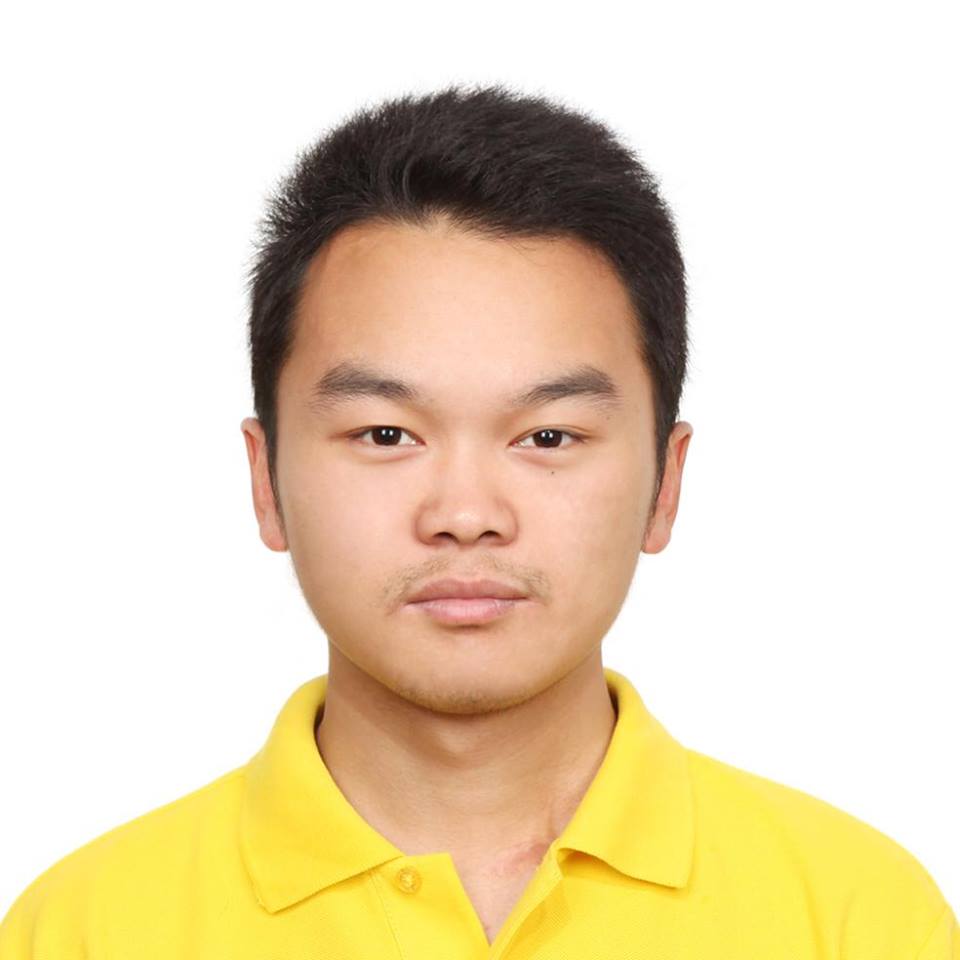}}]{Chuangchuang Sun}
is currently a postdoctoral associate in the department of aeronautics and astronautics at MIT. He received his Ph.D. in August 2018 from the Ohio State University and a B.S. degree from the Beijing University of Aeronautics and Astronautics, China in 2013, both in Aerospace Engineering. His research interests focus on control, optimization, reinforcement learning with applications in robotics and space systems. 
\end{IEEEbiography}

\begin{IEEEbiography}[{\includegraphics[trim=80 0 50 0,width=.9in,clip,keepaspectratio]{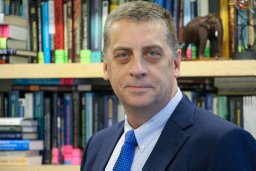}}]{Jonathan P. How} is the Richard C. Maclaurin Professor of Aeronautics and Astronautics at the Massachusetts Institute of Technology.  He received a B.A.Sc. (aerospace) from the University of Toronto in 1987, and his S.M. and Ph.D. in Aeronautics and Astronautics from MIT in 1990 and 1993, respectively, and then studied for 1.5 years at MIT as a postdoctoral associate. Prior to joining MIT in 2000, he was an assistant professor in the Department of Aeronautics and Astronautics at Stanford University.  Dr. How was the editor-in-chief of the IEEE Control Systems Magazine (2015-19) and is an associate editor for the AIAA Journal of Aerospace Information Systems and the IEEE Transactions on Neural Networks and Learning Systems. He was an area chair for International Joint Conference on Artificial Intelligence (2019) and will be the program vice-chair (tutorials) for the Conference on Decision and Control (2021).  He was elected to the Board of Governors of the IEEE Control System Society (CSS) in 2019 and is a member of the IEEE CSS Technical Committee on Aerospace Control and the Technical Committee on Intelligent Control. He is the Director of the Ford-MIT Alliance and was a member of the USAF Scientific Advisory Board (SAB) from 2014-17. His research focuses on robust planning and learning under uncertainty with an emphasis on multiagent systems, and he was the planning and control lead for the MIT DARPA Urban Challenge team.  His work has been recognized with multiple awards, including the 2020 AIAA Intelligent Systems Award. He is a Fellow of IEEE and AIAA.   
\end{IEEEbiography}

\EOD

\end{document}